\documentclass[10pt,twocolumn,english,aps,superscriptaddress]{revtex4-1}

\usepackage[T1]{fontenc}
\usepackage[latin9]{inputenc}
\usepackage{color}
\usepackage{babel}
\usepackage{latexsym}
\usepackage{amsmath}
\usepackage{amssymb}

\usepackage{graphicx}
\usepackage{times}   
\usepackage[unicode=true,pdfusetitle,bookmarks=false,colorlinks=true,citecolor=blue,urlcolor=blue,linkcolor=red]{hyperref}

\makeatletter

\@ifundefined{textcolor}{}
{%
 \definecolor{BLACK}{gray}{0}
 \definecolor{WHITE}{gray}{1}
 \definecolor{RED}{rgb}{1,0,0}
 \definecolor{GREEN}{rgb}{0,1,0}
 \definecolor{BLUE}{rgb}{0,0,1}
 \definecolor{CYAN}{cmyk}{1,0,0,0}
 \definecolor{MAGENTA}{cmyk}{0,1,0,0}
 \definecolor{YELLOW}{cmyk}{0,0,1,0}
}

\@ifundefined{date}{}{\date{}}
\AtBeginDocument{
  
}
\makeatother

\newcommand{\SAVE}[1]{}

\newcommand{\prlsec}[1]{\emph{#1---}}

\begin{document}
\renewcommand\abstractname{}

\title{Phase diagram of the Kondo lattice model on the Kagome lattice}
\author{Shivam Ghosh}
\affiliation{Laboratory of Atomic And Solid State Physics, Cornell University, Ithaca, NY 14853, USA}
\author{Patrick O' Brien}
\affiliation{Department of Physics, Binghamton University, Binghamton, NY}
\author{Christopher L. Henley}
\affiliation{Laboratory of Atomic And Solid State Physics, Cornell University, Ithaca, NY 14853, USA}
\author{Michael J. Lawler}
\affiliation{Department of Physics, Binghamton University, Binghamton, NY}

\date{March 24, 2015}

\begin{abstract}
We consider the potential for novel forms of magnetism arising from the subtle interplay between electrons and spins in the under-screened kagome Kondo lattice model. At weak coupling, we show that incommensurate non-coplanar multi-wave vector magnetic orders arise at nearly all fillings and that this results from Fermi surface effects that introduces competing interactions between the spins. At strong coupling, we find that such complex order survives near half filling despite the presence of ferromagnetism at all other fillings. We show this arises due to state selection among a massive degeneracy of states at infinite coupling. Finally, we show that at intermediate filling, only commensurate orders seem to survive. But these orders still include non-coplanar magnetism. So, the mere presence of both local moments and itinerant electrons enables complex orders to form unlike any currently observed in kagome materials.

\end{abstract}

\maketitle

\prlsec{Introduction} 
Two paths are known whereby local Hamiltonians in lattice models can stabilize complex spin order -- meaning both that the spin configurations are complex in space, and that the phase diagram contains a zoo of different phases. One well-known path to such complexity is \emph{state frustration}--meaning the ground states are massively degenerate. Any small perturbation, such as disorder\cite{shender}, dipolar interactions\cite{chalker}, or simply the intrinsic quantum or thermal fluctuations\cite{henley,chern0,Uzi}, then suffice to select a particular state as the unique ground state.

A second path to complexity is through \emph{frustrated interactions}, i.e. there are multiple kinds of  Heisenberg spin couplings that cannot be satisfied simultaneously.  Complexity may be realized with as few as {\it two} isotropic neighbor distances, but only when the spin sites form a {\it non-Bravais} lattice, such as Kagome or Pyrochlore lattices or when the interactions are non quadratic\cite{martinspiral,satorucubic}: rigorously, on Bravais lattices  with isotropic Heisenberg quadratic couplings--at most simple coplanar spin spirals are realized\cite{kaplan}. 

Stable {\it noncoplanar} complex spin states are particularly intriguing for their unusual rigid-body-like order parameters. They are also motivated experimentally as they realize an anomalous Hall effect due to Berry phases\cite{martin,nagaosa,Taillefumier,akagi}, and theoretically since if such a phase loses long range order at sufficiently small spin-length, it is expected to become a {\it chiral spin liquid}, induced without any spin-orbit effects\cite{messio}.

Even more complex behavior is possible when the frustrated spin-spin interactions decay slowly with distance.That is easily realized by coupling local moments to a band of {\it fermions}, which mediate oscillating couplings between the local Heisenberg moments -- the so called Kondo Lattice Model (KLM)\cite{martin, akagi,akagi1,chern}.

In this letter, we show that at weak coupling, the kagome KLM supports incommensurate, non-coplanar, multi-wave vector spin ordering. This motivated us to seek the stability of these phases at intermediate and strong coupling and understand the potential for such novel magnetism to be discovered in materials. We show, by extending recipes to identify and classify states laid out previously\cite{sklan and matt}, that these novel complex orders arise from competing interactions introduced by the fermions near their Fermi surface. It therefore appears to be a weak coupling phenomena. At strong coupling, however, we discover such complex phases survive near half filling due to a separate mechanism: state selection from a massive degeneracy of states at infinite coupling. We then turn to intermediate coupling, where both state selection and competing iteractions are presumably at play, but find that dominantly commensurate orders appear to survive. But even here, some of the orders are non-coplanar so that at any coupling, complex order beyond any form currently observed in kagome materials are possible.

\prlsec{Complex orders at weak coupling} 
We adopt the KLM Hamiltonian given by
\begin{equation}
\mathcal{H}_{KLM}=-t\sum_{\langle ij\rangle,\alpha,\beta}c_{i(\alpha)\sigma}^{\dagger}c_{j(\beta)\sigma}-J_{K}\sum_{i=1,\alpha}^{N}\mathbf{S}_{i(\alpha)}\cdot \boldsymbol{s}_{i(\alpha)} 
\label{eq:hamiltonian}
\end{equation}
The first term is nearest-neighbor hopping with amplitude $t$ of a single band of noninteracting electrons, with creation operator $c_{i(\alpha)\sigma}^{\dagger}$ at unit cell $i$ and sublattice $\alpha$. The second term is the Kondo coupling, with $\boldsymbol{s}_{i(\alpha)}$ being the electron spin and $\mathbf{S}_{i(\alpha)}$ being classical Heisenberg spins representing the local moments. We seek the ground state configuration of the local moments $\{\mathbf{S}_{i(\alpha)}^{opt}\}$, for every fermion filling and coupling $J_K$ on the premise that we will find complex orders and a complex phase diagram.

Previous methods for finding $\{\mathbf{S}_{i(\alpha)}^{opt}\}$, were either variational MC, with costly fermionic diagonalization at each MC step (this has been recently overcome by an efficient algorithm \cite{barros,kato} allowing exploration of large system sizes) or else a variational optimization of $\langle \mathcal{H}_{KLM}\rangle$ using a trial basis~\cite{akagi}, limited to commensurate orders with small unit cells. 

We instead will study the limits $J_{K}/t \ll 1$ and $J_K/t \gg 1$ perturbatively. This allows access to large system sizes even with incommensurate orders. We begin here with the $J_{K}/t \ll 1$ limit but will see that extending it with the variational approach will motivate a study of the opposite limit. The effective Hamiltontian in the $J_{K}/t \ll 1$ regime is of the form:
\begin{multline}
\mathcal{H}_{eff} =\dfrac{1}{2}\sum_{i(\alpha),j(\beta)}J_{i(\alpha)j(\beta)}\mathbf{S}_{i(\alpha)}\cdot \mathbf{S}_{j(\beta)} + \\
	\dfrac{1}{4!}\sum_{i(\alpha),\ldots,l(\delta)}\hspace{-0.1in} K_{i(\alpha),j(\beta),k(\gamma),l(\delta))}
	\mathbf{S}_{i(\alpha)}\cdot \mathbf{S}_{j(\beta)} \mathbf{S}_{k(\gamma)}\cdot \mathbf{S}_{l(\delta)} + \ldots 
\label{eq:rkky}
\end{multline}
In the limit $J_K \to 0$ we need only focus on the first term with the RKKY couplings $J_{i(\alpha)j(\beta)}$. To compute these couplings at $T=0$, we take a grid in reciprocal space; the corresponding lattice sizes were up to $N=3 \times 36^{2}$. We first locate the Fermi surface corresponding to the chosen filling, then numerically evaluate the usual analytic formula for $J_{i(\alpha)j(\beta)}$ from second-order perturbation theory (see $\mathbf{S.I.}$ I)\footnote{bcaveraging}.

It is well known that RKKY interactions introduce frustrated interactions among the spins. The $J_{i(\alpha)j(\beta)}$ here are no different in principle, but the degree of this frustration, shown in the low temperature spin order $\{\mathbf{S}_{i(\alpha)}^{opt}\}$ obtained from zero and finite temperature Monte Carlo(MC), is striking. 
For example, at $n=0.325$, near $1/3$ filling, we find a twisted $\sqrt{3}\times\sqrt{3}$ state with an incommensurate wave vector and slight non-coplanarity. This state smoothly evolves to the coplanar $\sqrt{3} \times \sqrt{3}$ order at $n=1/3$. As $n$ approaches $n=5/12$, we find the "Cuboc1"\cite{messio} state that is commensurate but non-coplanar with 12 different spins in its unit cell all pointing to the edges of a cube. But these are among the simplest states we found. At $n=0.488$, for example, we found a three wave vector, incommensurate and non-coplanar state. 

\begin{figure}[htpb]
\centering
\includegraphics[width=\linewidth]{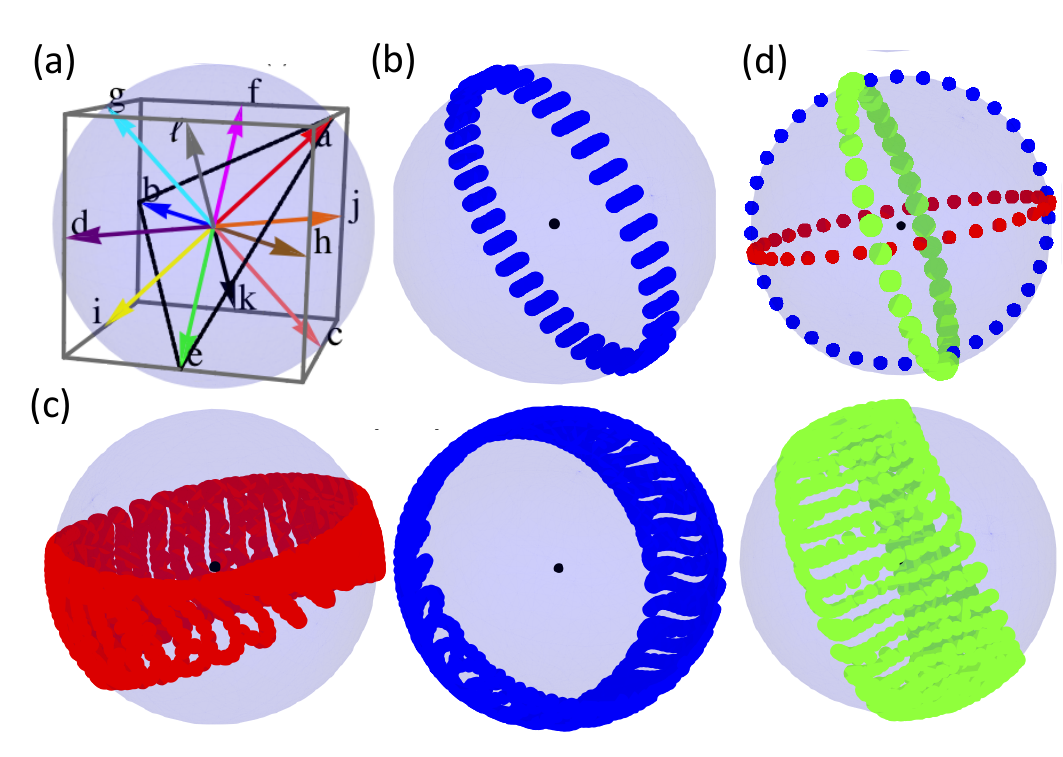} \caption{Some ground state spin orders in of the kagome Kondo model at small $J_K/t$. (a) The "Cuboc1" state\cite{messio1}, found at $n=5/12$, consisting of 12 different spins pointing to the edges of a cube. (b) An incommensurate twisted $\sqrt{3}\times\sqrt{3}$ state found at $n=0.325$ (c) Sublattice separated common origin plots of a complex state found at $n=0.488$ on an $N=3\times 36^2$ cluster. (d) A simplified version of the three common origin plots presented in (c) emphasizing the 90$^o$ rotational symmetry of the respective planes.}
\label{fig:MC}
\end{figure}

To draw these conclusions, we have plotted the spins with the tail of each spin vector at a common origin\cite{sklan and matt} as shown in Fig.~\ref{fig:MC} and studied their structure factor. The common origin plot of the "Cuboc1" state is shown in Fig.~\ref{fig:MC}(a) and has a cuboctahedral structure. Similarly, the coplanar nature of the twisted $\sqrt{3}\times\sqrt{3}$ state is readily apparent as shown in Fig.~\ref{fig:MC}(b). However, the complex state at $n=0.488$ would appear nearly incomprehensible in a common origin plot, having spins that point in nearly all directions. But constructing the common origin plots for each sublattice separately, as shown in Fig.~\ref{fig:MC}(c), reveals a different dominant plane for each of them. These planes cross each other at 90$^0$ angles as shown in the simplified structure plotted in Fig.~\ref{fig:MC}(d). In this way the common origin plots reveal a simple structure of many of the complex states discovered by our MC calculations. 

To identify the wave vector dependence of the spin order, we have also computed the sublattice dependent structure factor:
\begin{equation}
 \eta_{\alpha}(\mathbf{q})=\frac{|\mathbf{S}_{\alpha}^{opt}(\mathbf{q})|^{2}}{\sum_{\mathbf{q}\in 1st B.Z.} |\mathbf{S}_{\alpha}^{opt}(\mathbf{q})|^{2}}.
\label{eq:eta}
\end{equation}
Here $\mathbf{S}_{\alpha}^{opt}(\mathbf{q})$ is the Fourier transform of $\{\mathbf{S}_{i(\alpha)}^{opt}\}$. This was needed, for example, to see that the twisted $\sqrt{3}\times\sqrt{3}$ state indeed had a wave vector near the Brillouin zone corner (the $\mathbf{K}$ point) and that all sublattices had the same wave vector. Conversely, it reveals that each sublattice of the state found at $n=0.488$ had a different wave vector but all three wave vectors were related by $2\pi/6$ rotations. The main features of the spin orders are thus revealed in this structure factor and the common origin plots.

So, the Monte-Carlo results demonstrate that frustrated interactions in this model drive both unusually complex magnetism and a complex phase diagram.

\prlsec{Luttinger-Tisza and the Fermi surface} 
To understand how the competing interactions drive the wave vector composition, sublattice dependence, incommensurate order and non-coplanarity, we have carried out a Luttinger-Tisza (L. T.) analysis\cite{lt}. The method begins by diagonalizing $J_{\alpha \beta}(\mathbf{q})$, the spatial Fourier transform of $J_{i(\alpha),j(\beta)}$:
\begin{equation}
J_{\alpha \beta}(\mathbf{q})u^{\nu}(\mathbf{q})=\lambda^{\nu}(\mathbf{q})u^{\nu}(\mathbf{q})
\label{eq:lt}
\end{equation}
where the $\nu$ index label three bands in the sublattice space and $\lambda^{\nu}(\mathbf{q})$ is the eigenvalue corresponding to the eigenstate $u^{\nu}(\mathbf{q})$.  Then we determine the symmetry related wave vectors $\{\mathbf{Q}_{L.T.}\}$ with the lowest eigenvalues of $J_{\alpha \beta}(\mathbf{q})$ and the corresponding eigenvectors $u^{\nu}(\mathbf{Q})_{L.T.}$. This information should predict the main features of the ordering patterns found numerically.

\begin{figure}[htpb]
\centering
\includegraphics[width=\linewidth]{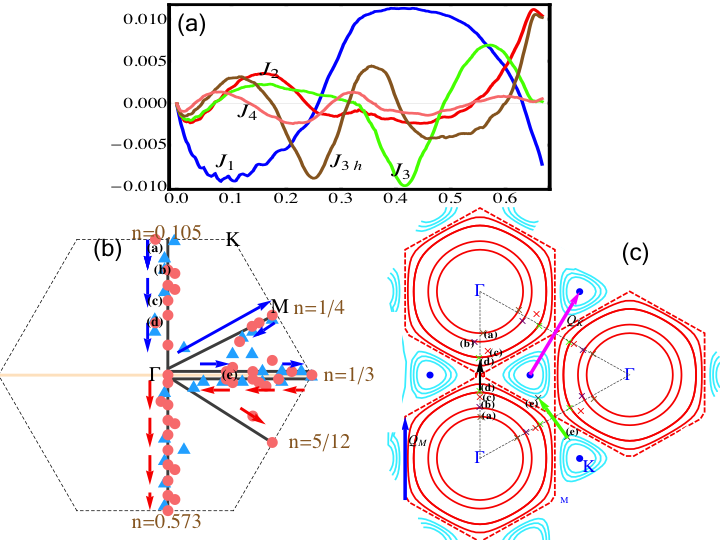} \caption{L. T. analysis of the couplings $J_{i(\alpha),j(\beta)}$. (a) The nearest, $J_1$, next nearest, $J_2$ and two third nearest $J_3$, $J_{3h}$ neighbor couplings plotted as a function of filling $n$. (b) The trajectory of the predicted ordering wave vector $\mathbf{Q}_{L.T.}(n)$ as $n$ is varied for lattice sizes $N=3\times 24^{2}$ (triangles) and $N=3\times 36^{2}$ (circles). (c) Comparison between $\mathbf{Q}_{L.T.}$ and the Fermi surface. Red contours show $0<n<1/4$, dashed red line is the F.S. at $n=1/4$ and blue contours show $1/4 < n < 1/3$. Several $\mathbf{Q}_{L.T.}(n)$ are labeled as arrows. The black arrow is near $n=0.488=2/3-0.179$, the blue arrow is at $n=1/4$, where $\mathbf{Q}_{L.T.}(n)=\mathbf{Q}_M$, and the green arrow is near $n=0.325$.}
\label{fig:LT}
\end{figure}

To get a sense of what we might find, we have plotted the first five couplings as a function of filling in Fig.~\ref{fig:LT} (a). Here, negative (positive) couplings correspond to ferromagnetic (antiferromagnetic) interactions. Clearly, at $n=0$ we should find ferromagnetism. For $n\lesssim 0.2$ we find a dominant ferromagnetic $J_1$ coupling and small $J_2$, $J_3$, $J_{3h}$ and $J_4$ couplings. This suggests the ferromagnetic state should become a spiral. Near $n=1/3$ we expect a $\sqrt{3}\times\sqrt{3}$ coplanar state known to exist\cite{chubukov} at positive $J_1$, with small positive $J_3$, $J_{3h}$. A similar argument leads to the $q=0$ state near $n=0.5$. However, at all other fillings simple arguments such as these are not enough to estimate what state will minimize the energy.

Plotting the wave vectors $\mathbf{Q}_{L.T.}$, as shown in Fig.~\ref{fig:LT}(b) reveals that indeed the predictions from the L. T. method are more complex. The wave vectors do begin at the $\mathbf{Q}_\Gamma=0$ point at $n=0$ but they then march towards the Brillouin zone boundary at the $\mathbf{M}$ point by $n=0.105$ and returns to the $\Gamma$ point as $n$ approaches $1/4$. However, exactly at $n=1/4$, and probably for a small interval on either side of it, $\mathbf{Q}_{L.T.}(n)$ jumps back to the $\mathbf{M}$ point -- a first order transition. For $1/4\lesssim n<1/3$, the $\mathbf{Q}_{L.T.}(n)$ resumes moving smoothly from the $\Gamma$ point, but switches trajectory moving towards the $\mathbf{K}$ point and reaching it uneventfully at $n=1/3$. For $1/3 < n < 2/3$ wave vectors are the same as those at $2/3 - n$. 

This complex dance of the wave vector must be related to the evolution of the $J_K =0$ Fermi surface. To see more specifically how the two are related, we have plotted in Fig.~\ref{fig:LT}(c) several $\mathbf{Q}_{L.T.}$ on top of this Fermi surface as connecting vectors. We see from this plot that it is the points of maximum curvature of the surface that determine $\mathbf{Q}_{L.T.}$. This is true at both $n=0.325$ where we found the twisted $\sqrt{3}\times\sqrt{3}$ state in MC and at $n=0.488 = 2/3 - 0.179$ where we found the multi-wave vector state.  An exception to this rule occurs near $n=5/12$, where the wave vector nests the flat regions of this Fermi surface. Here we found the cuboc1 state in MC. The van Hove singularity at $n=1/4$ also explains the sudden jump in the wave vector as $n$ approaches this filling. So the wave vector predictions from the L. T. analysis are entirely determined by the geometry of the Fermi surface. A similar observation was also noted in a recent study of KLM on Bravais lattices\cite{satoru}.

Finally, we look at both $Q_{L.T.}$ and the eigenvectors $u^{\nu}(\mathbf{Q}_{L.T.})$ to compare with the MC results. At $n=1/3$ we find $Q_{L.T.}$ is at the $K$ point which is the wave vector of the $\sqrt{3}\times\sqrt{3}$ state and $u^{\nu}(\mathbf{Q}_{L.T.}) = \sqrt{1/3}(1,1,1)$ (equal weights on all sublattices). This is in agreement with both the common origin plots and the structure factor $\eta_{\alpha}(\mathbf{Q}_{L.T.})$ obtained in MC. At $n=5/12$, we find $u^{\nu}(\mathbf{Q}_{L.T.}) = \sqrt{1/2}(1,1,0)$ (for the other two $\mathbf{Q}_{L.T.}$, related by $2\pi/3$ rotations, the vectors are $\sqrt{1/2}$(1,0,1) and $\sqrt{1/2}$(0,1,1)) At $n=0.488$ we find $u^{\nu}(\mathbf{Q}_{L.T.})$ has weight on only one sublattice for a given $\mathbf{Q}_{L.T.}$, so $u^{\nu}(\mathbf{Q}_{L.T.}) \approx (1,0,0)$. Both this relationship between $\mathbf{Q}_{L.T.}$ and the eigen-vector $u^{\nu}(\mathbf{Q}_{L.T.})$ is precisely that found in $\eta_{\alpha}(\mathbf{q})$. So the L. T. method both predicts the wave vector and sublattice depedence of the spin order $\{\mathbf{S}_{i(\alpha)}^{opt}\}$ we found in MC.

Thus, the Luttinger-Tisza method is in simple agreement with our MC calculations and reveals that Fermi surface geometry dictates the incommensuratation of the complex ordering patterns in the limit $J_K/t\to0$.

\prlsec{Phase diagram}
With the knowledge discussed above, it seems like we can now find the phase diagram in the limit $J_K/t \ll 1$.  But since Fermi surface geometry dictates ordering wave vectors, every filling $n$ will have a different ground state spin configuration. In group theory language, the spin order $\{\mathbf{S}_{i(\alpha)}^{opt}\}$ will be built out of different representations of $SO(3) \times G_{kagome}$ where $SO(3)$ is the classical spin group and $G_{kagome}$ is the kagome space group. So every spin configuration likely belongs to a different phase.

To navigate this issue, here we will adopt a pragmatic approach to depict the phase diagram.  We first label a phase by a triplet of symbols that characterize the wave vector dependence of the spin pattern on each of the three sub-lattices. For example, we label the spin configuration presented in Fig.~\ref{fig:MC}(b), as the $(a_{1},a_{1},a_{1})$ phase. The symbol $(a_{1},a_{1},a_{1})$ means that all three sublattices have the same wave vector $\mathbf{Q}_a=2\pi(7/24,0) \approx \mathbf{K}$. The subscript $1$ denotes that they have the same rank in the list of sublattice dependent structure factor values $\eta_\alpha(\mathbf{q})$ sorted with largest magnitude first. The state at $n=0.488$ we label as the $(a_{1},b_{1},c_{1})$ phase because each sublattice has a different wave vector: $\mathbf{Q}_a$, $\mathbf{Q}_b$, $\mathbf{Q}_c$. But these are related by $2\pi/3$ rotations and have equal Fourier weights ($\eta_{1}(\mathbf{Q}_a)=\eta_{2}(\mathbf{Q}_b)=\eta_{3}(\mathbf{Q}_c)$). These labels coarse grain the phase diagram.

We then identify two spin configurations as belonging to the same phase family if they share the same broken symmetries and their labels evolve smoothly with $n$ (for an example of a family of spin configurations evolving smoothly with filling, see S.I. II). This approach is incomplete\cite{fn:statelabel}
but useful because two spin configurations that violate this rule are certainly in different phases.

\begin{figure}[htpb]
\centering
\includegraphics[width=\linewidth]{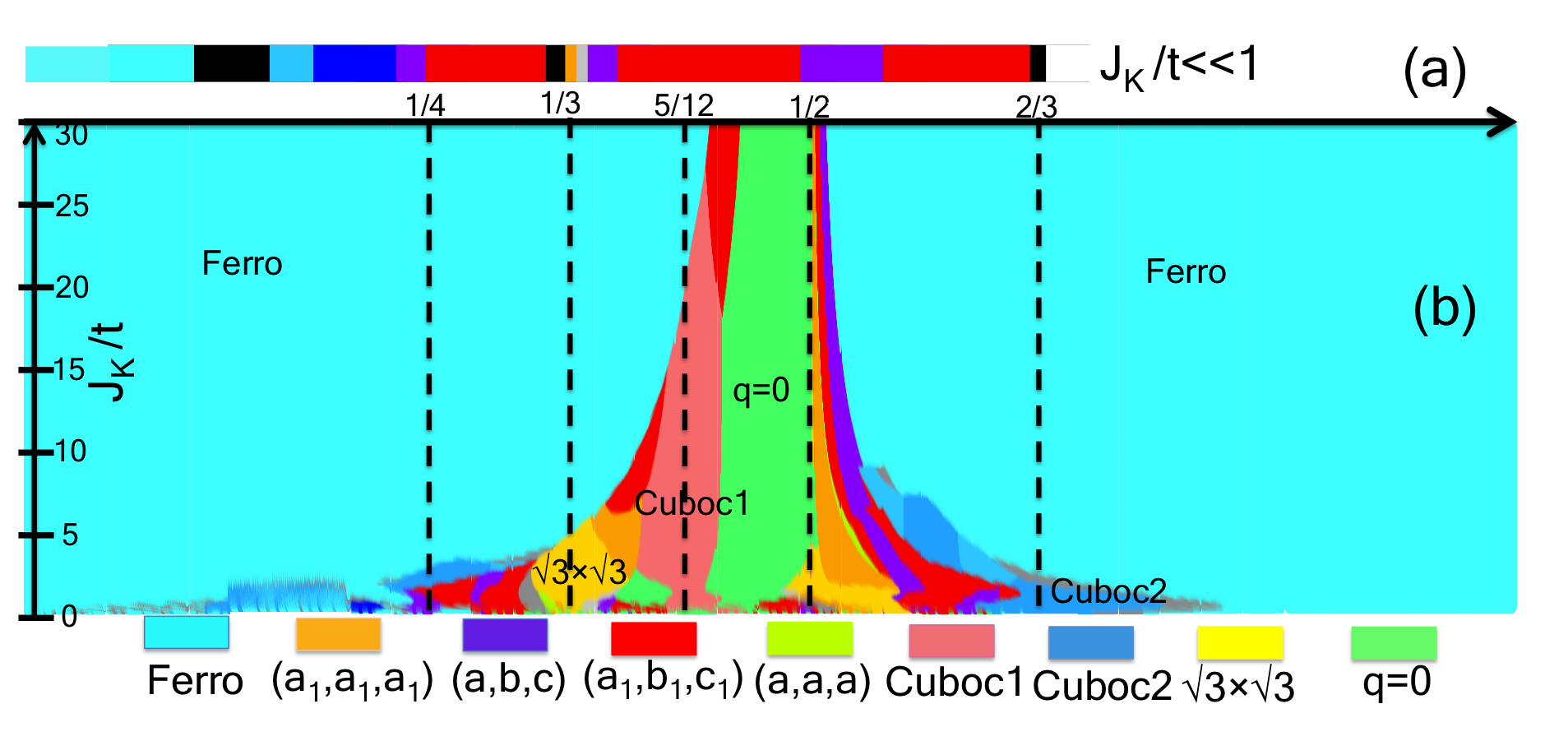} \caption{Phase diagram of the KLM (Eq.  \eqref{eq:hamiltonian}) . (a) exact phase diagram in the limit $J_K/t \ll 1$. Here black regions denote fillings where multiple phases were competitive and our approach could not identify a single dominant phase (b) Variational phase diagram showing different competing phases at finite $J_K/t$. The range along the horizontal filling axis is $n=(0,1)$ at unit intervals and along coupling axis is $J_{K}=[0.1,10]$ on a grid of $0.05$. Energies at every point $(n,J_{K})$ in the phase diagram are averaged over 100 values of the boundary phases.}
\label{fig:PD}
\end{figure}

Using this approach, we construct the phase diagram in Fig. \ref{fig:PD}(a) by computing in MC the spin configurations at each filling in the range $0 \leq n \leq 2/3$ (at $n>2/3$ we start filling a flat band and the approach fails). This leads to order $N\approx 3\times 36^{2}$ different spin configurations (about the number of sites in the lattice) but our labeling approach groups them into eight phases (see Supplementary Information S.I. III and IV). Five of these are the well known commensurate phases ferromagnetic, $cuboc1$,$cuboc2$, the $q=0$ and $\sqrt{3}\times\sqrt{3}$, and the rest are incommensurate phases.

Given the large database of energetically competitive spin configurations obtained from our numerics, and that the energy of each at finite $J_K/t$ is easily computed, we can also construct a variational phase diagram away from $J_K/t \ll 1$. In computing it, we additionally add the cuboc2 state\cite{domenge}. The result is presented in Fig.~\ref{fig:PD}. 

We see several new features in the phase diagram as $J_K/t$ approaches 1 and larger values. Ferromagnetism begins to dominate much of it above $J_K/t=1$. The complex ordered states stable at small $J_K/t$ fan out at first but then most vanish and commensurate orders dominate the intermediate $J_K/t$ regime.  
This provides further evidence that complex orders at small $J_K/t$ resulted primarily from Fermi surface effects.  However, there are exceptions to this rule. Some incommensurate orders re-emerge over sizable regions at larger $J_K/t$ especially near $n=0.5$. So the phase diagram remains complex at finite $J_K/t$ but ferromagnetism begins to dominate much of it.

\prlsec{State selection at $J_K/t$}
The dominance of ferromagnetism in the phase diagram of Fig.~\ref{fig:PD} provokes the question of whether there is a threshold value of $J_K/t$ above which it is stabilized at all fillings. To see if this happens, here we consider the $J_K/t \gg 1$ ``double-exchange model'' regime. 

As is well known\cite{anderson1955,degennes1960}, in the $J_K/t \gg 1$ limit we can switch to coordinates with local quantization axes pointing along the $\vec S_{i(\alpha)}$ classical spin directions and find that the energy bands separate into the spin ``up" bands and spin ``down" bands with a gap proportional to $J_K$. For $n<1/2$, we can then integrate out all the higher energy ``down'' bands and obtain an effective model for the up bands
\begin{equation}
H_{up} = \sum_{\langle i(\alpha),j(\beta)\rangle} t\cos(\theta_{i(\alpha),j(\beta)}/2)e^{ia_{i(\alpha)j(\beta)}} u^\dagger_{i(\alpha)} u_{j(\beta)} + h.c.
\end{equation}
where spin-less fermion operator $u^\dagger_{i(\alpha)}$ creates an ``up" spin on site $i(\alpha)$, $\theta_{i(\alpha),j(\beta)}$ is the relative angle between classical spin vectors on the two sites and $a_{i(\alpha)j(\beta)}$ is a vector gauge potential arising from the non-coplanarity of spins\cite{nagaosa}.

This hopping process then contributes substantially to the energy if the spins are parallel and $\theta_{i(\alpha),j(\beta)}=0$ but contributes nothing if the spins are anti-parallel with $\theta_{i(\alpha),j(\beta)}=\pi$.  So in total, there is a large kinetic energy gain if the spins all point parallel and any deviation from this inhibits the electrons from gaining kinetic energy. Hence, in the $J_K/t \gg 1$ regime, ferromagnetism will likely dominate and the complex orders discovered at small $J_K/t$ will vanish. However, this argument fails at $n=1/2$ since here we completely fill the up bands and hopping costs energy of order $\sim J_{K}$.

To study the breakdown of this argument, we have carried out a perturbative calculation to second order in $t/J_K$ at $n=1/2$. It reveals nothing but the nearest neighbor anti-ferromagnetic Heisenberg model $J\sum_{\langle i(\alpha),j(\beta)\rangle}\mathbf{S}_{i(\alpha)}\cdot\mathbf{S}_{j(\beta)}$ with $J = t^2/J_K$. So at $n=1/2$, antiferromagnetic states dominate over the Ferromagnetic state even at $J_K/t \to\infty$.

Achieving a full understanding of the $n\approx 1/2$, large $J_K/t$ regime is therefore a {\emph state selection problem}. It is akin to the state selection through order by disorder by either temperature or quantum fluctuations of the nearest neighbor Kagome Heisenberg model. The fluctuations that produce the selection here, however, results from the fermions as they try to hop around in the presence of the spins. 

In the presence of temperature\cite{zhitomirsky,chern3} or quantum\cite{chubukov,chan} fluctuations, it is known that the $\sqrt{3}\times\sqrt{3}$ state is selected. It seems natural to expect that this state will also appear $n=1/2$. But, according to our variational calculations, this is not the case. The cuboc1 state wins over the other two states (Cuboc1 is also selected in a more trivial state selection problem in the presence of a nearest neighbor super-exchange interaction, see S.I. IV). 

Motivated by this switch from ferromagnetism to antiferromagnetism, we have produced a variational phase diagram in the vicinity of $n=1/2$ up to $J_K/t=10^4$. We see in this plot that even at these extreme values, the antiferromagnetic state at $n=1/2$ does not make a direct transition into the ferromagnetic state. Instead, a series of commensurate and incommensurate orders intervene.


So, state selection among all possible magnetic states, both ferromagnetic and antiferromagnetic, is responsible for the survival of complex order near $n=1/2$ and $J_k/t\gg 1$.

\begin{figure}[htpb]
\centering
\includegraphics[width=\linewidth]{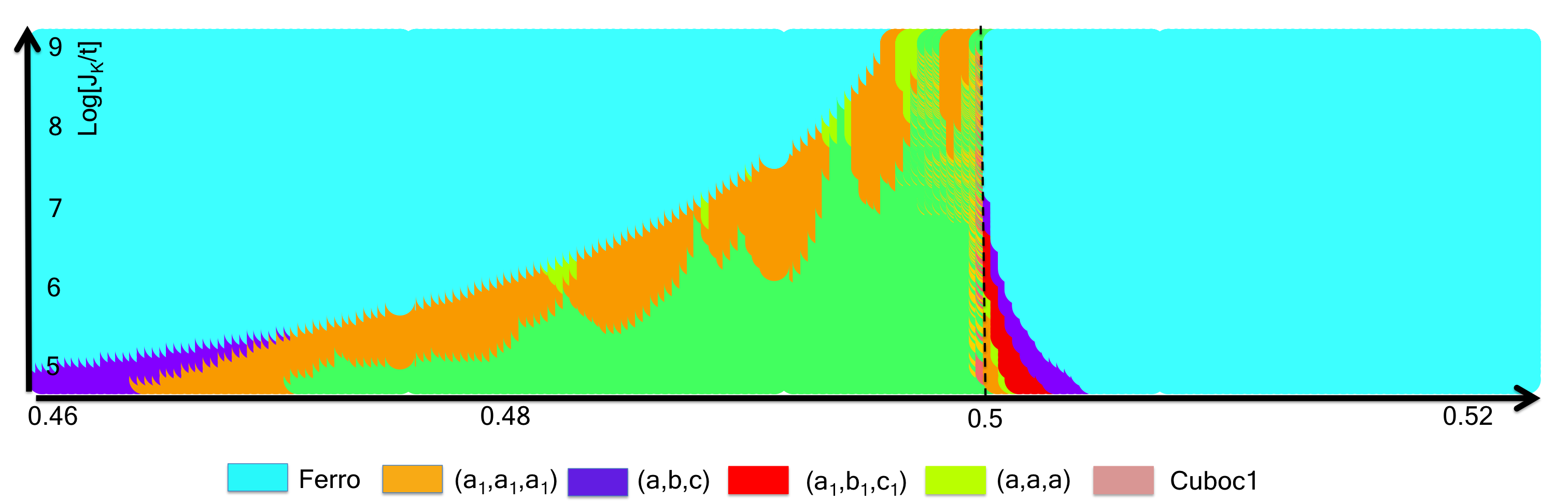} \caption{Variational phase diagram in the large-$J_K/t$ regime. This phase diagram includes 48 states taken from the set of MC ground states at $J_k/t\ll1$, and the set of commensurate orders consisting of the $q=0$, $\sqrt{3}\times\sqrt{3}$, cuboc1,  cuboc2 and ferromagnetic state. Notice, the dominance of the ferromagnetic state throughout the phase diagram except near $n=1/2$ where antiferromagnetic states survive including the $(a_1,b_1,c_1)$ state shown in Fig. \ref{fig:MC}(d) that is non-coplanar and incommensurate.}
\label{fig:largeJK}
\end{figure}
 
\prlsec{Search for complex magnetic order} 
Our results show that the under screened Kondo regime supports complex forms of magnetic order with non-coplanar, incommensurate, and multi-wave vector properties throughout much of its phase diagram and therefore provide useful insight that can guide the search for such magnetism in materials. The most exotic form of this magnetism will likely be found either near $n=1/2$ at large $J_K/t$ supported by a state selection mechanism or for all $n$ at small $J_K/t$ due to F.S. effects. However, novel commensurate non-coplanar states are still possible in the intermediate regime and could be discovered for example by studying a range of doping.  

Real materials, however, will unlikely be described by the idealized kagome Kondo model we study. Likely, their band structure will be much more complicated and three dimensionality may be important. The Luttinger-Tisza method, MC calculations at small $J_K/t$ and hopping model at large $J_K/t$ are easily extended to these cases. Further, additional physics, such as impurities, additional spin-spin interactions, various magnetic anisotropies can also be handled by these methods. So the methods we establish here will likely play a role in the search and study of complex forms magnetism arising from under screened Kondo physics.

Our approach to the variational problem may also prove useful in the study of real materials. By creating a database of incommensurate non-coplanar states stabilized at small $J_K/t$, we discovered that commensurate states, though prominent near $J_K/t=1$, can lose out to ``reentrant'' incommensurate orders that appear at different fillings than they were found in the limit of small $J_K/t$ regime.  Such orders were neglected in previous studies\cite{akagi} of KLM phase diagram.  

Finally, we would like to stress that non-coplanar orders appear rarely in Kagome systems. Our results suggest that, in Kondo-coupled systems, such as potentially the layered itinerant Kagome ferromagnet $\mbox{Fe}_{3}\mbox{Sn}_{2}$\cite{fenner} and doped $\mbox{FeCrAs}$\cite{julian} with small spin-orbit effects, non-coplanar commensurate and incommensurate orders are abundant.

\prlsec{Acknowledgement} We would like to thank C. D. Batista, J. Chalker, M. Gingras, I. Martin, Y. Motome and especially G.W. Chern for discussions. CLH and SG acknowledge support from NSF grant NSF DMR-1005466.


\newpage
\section{Supplementary Information S.I. I}
\label{sec:S1}

We will derive the RKKY couplings via two different methods in this section and compare results obtained from each method. The first way to get $J_{i(\alpha) j(\beta)}$ is in Fourier space using second order perturbation theory. We define the following electron operators in momentum space:

\begin{equation}
c_{\mathbf{k}}^{\alpha}=\frac{1}{\sqrt{N/3}}\sum_{i}c_{i(\alpha)}e^{i \mathbf{k}.\mathbf{R}_{i}}
\label{eq:ck}
\end{equation}

where $N/3$ is the number of unit cells and $\alpha$ is the sublattice index. Using Eq.\eqref{eq:ck} we first write down the free fermion part of the KLM Hamiltonian in the basis $| \mathbf{k}^{\alpha} \rangle$ as follows,

\begin{equation}
\left (\begin {array}{ccc}
0 & -2t \cos(\mathbf{k} \cdot \mathbf{a}_{12}) & -2t \cos(\mathbf{k}\cdot \mathbf{a}_{13}) \\
 -2t \cos(\mathbf{k} \cdot \mathbf{a}_{12})  & 0 & -2t \cos(\mathbf{k}\cdot \mathbf{a}_{23}) \\
 -2t \cos(\mathbf{k}\cdot \mathbf{a}_{13}) & -2t \cos(\mathbf{k}\cdot \mathbf{a}_{23}) & 0 \\
\end{array} \right)
\label{eq:h0}
\end{equation}

where $\mathbf{a}_{\alpha \beta}=\mathbf{a}_{\alpha}-\mathbf{a}_{\beta}$, $\mathbf{a}1=(0,0)$, $\mathbf{a}_{2}=(1,0)$ and $\mathbf{a}_{3}=(1/2,\sqrt{3}/2)$. Both $\mathbf{a}_{2(3)}$ are one half of lattice vectors that define the Kagome unit cell. Diagonalizing \eqref{eq:h0} yields three bands $\nu$ with single particle energies $\varepsilon_{\mathbf{k}}^{\nu}$, eigenvectors $u_{\mathbf{k},\alpha}^{\nu}$ and a new set of operators $c_{\mathbf{k}}^{\nu \dagger}=\sum_{\alpha} u_{\mathbf{k},\alpha}^{\nu}c_{\mathbf{k}}^{\alpha \dagger}$. We can now look at the Kondo perturbation to the free fermion dispersion. 

The Kondo part of Eq.1in \cite{maintext} can be expressed in real space as
\begin{eqnarray} \nonumber \mathcal{H}_{Kondo} =-\frac{J_{K}}{2} \sum_{i}\sum_{\alpha}(S_{i(\alpha)}^{-}c_{i(\alpha)\uparrow}^{\dagger}c_{i(\alpha) \downarrow}+ \\
S_{i(\alpha)}^{+}c_{i(\alpha)\downarrow}^{\dagger}c_{i(\alpha) \uparrow}+S_{i(\alpha)}^{z}c_{i(\alpha)\uparrow}^{\dagger}c_{i(\alpha) \uparrow}-S_{i(\alpha)}^{z}c_{i(\alpha)\downarrow}^{\dagger}c_{i(\alpha) \downarrow} )
\label{eq:hkondo}
\end{eqnarray}
 
 Second order perturbation theory will carry two copies of \eqref{eq:hkondo}, each of them sandwiched between pairs of electronic states inside $| \mathbf{k}_{in}^{\nu}  \rangle$ and outside $\mathbf{k}_{out}^{\nu'}$ the F.S. given by
 
 \begin{equation}
 E_{2}(n)=J_{K}^{2}\sum_{\mathbf{k}_{in},\mathbf{k}_{out},\nu,\nu'} \frac{|\langle \mathbf{k}_{in}^{\nu}| \mathcal{H}_{Kondo}| \mathbf{k}_{out}^{\nu'} \rangle|^{2}}{(\varepsilon_{\mathbf{k}_{in}}^{\nu}-\varepsilon_{\mathbf{k}_{out}}^{\nu'} )  }
 \label{eq:E2}
 \end{equation} 

Insertion of \eqref{eq:hkondo}  in to \eqref{eq:E2} produces 16 terms, out of which only 4 are non zero due to spin rotational invariance. Each of these four terms have the same contribution and $E_{2}(n)$, expressed in Fourier space, then becomes

\begin{equation}
E_{2}(n)=\sum_{\mathbf{q}\in 1st B.Z.} J_{\alpha \beta}(\mathbf{q}) \mathbf{S}_{\alpha}(\mathbf{q})\cdot \mathbf{S}_{\beta}(-\mathbf{q})
\end{equation}

where $J_{\alpha \beta}(\mathbf{q})$ is a $3 \times 3$ matrix in the sublattice basis and is given by

\begin{equation}
J_{\alpha \beta}(\mathbf{q})=-\frac{J_{K}^{2}}{2}\sum_{\mathbf{k}_{in},\nu, \nu'}\frac{  u_{\mathbf{k}_{in},\alpha}^{\nu*}  u_{\mathbf{k}_{in}+\mathbf{q},\alpha}^{\nu'}   u_{\mathbf{k}_{in}+\mathbf{q},\beta}^{\nu' *}  u_{\mathbf{k}_{in},\beta}^{\nu}   }{(\varepsilon_{\mathbf{k}_{in}+\mathbf{q}}^{\nu '}-\varepsilon_{ \mathbf{k}_{in} }^{\nu} ) }e^{- i \mathbf{q} \cdot \mathbf{a}_{\alpha \beta} }
\label{eq:jab}
\end{equation}

where $u_{\mathbf{q},\alpha}^{\nu}$ is an amplitude for destroying an electron with wave vector $\mathbf{q}$ in band $\nu$ and the summation is restricted to states $\mathbf{k}_{in}$ below the F.S. Note that in going from \eqref{eq:E2} to \eqref{eq:jab} we have switched dummy momentum indices from $(\mathbf{k}_{in},\mathbf{k}_{out})$ to $\mathbf{k}_{in},\mathbf{q}$. Use of zone symmetries ($C_{6}$ and mirror reflections) requires us to compute $J_{\alpha \beta}(\mathbf{q})$ for only $1/12$ of the zone. The real space couplings are obtained by inverse Fourier transforming \eqref{eq:jab}

\begin{equation}
J_{i(\alpha) j(\beta)}= \sum_{\mathbf{q}} J_{\alpha \beta}(\mathbf{q})e^{-i \mathbf{q} \cdot \mathbf{R}_{ij}}
\label{eq:jij}
\end{equation}
 
 There are two limitations of computing the set of $\{J_{i(\alpha)j(\beta)}\}$ via the method above. Firstly, symmetries in the zone at any filling lead to degeneracies in the single particle energies (typically six leading to twelve missing electronic states) requiring us to "hop" over fillings so as to avoid zero energy denominators in \eqref{eq:jab}. The second limitation is a more severe form of the first constraint, where, for a window of fillings near Van Hove points, computation of \eqref{eq:jab} is restricted by large parallel parts of the F.S. with degenerate energies. The first limitation is resolved by averaging \eqref{eq:jab} over two non-local boundary phases which break the six fold symmetry to two (associated with spins). To circumvent the second restriction we propose an alternative methodology for computations of $\{ J_{i(\alpha)j(\beta)} \}$ as follows.
 
 The second method for extracting the set of couplings is an approach in real space, cruder in spirit, but works as well as the explicit calculation in Fourier space. Using exact numerical diagonalization, we evaluate the single particle energies of fermions in Eq.1 in \cite{maintext} in the background of a set of random spin configurations of size $N_{s}$ for a given $J_{K}$. These single particle energies are summed up to the F.S. to find the total energy $E_{ED}(n)$ as a function of filling. Each element from the set of $\{E_{ED}(n)\}$ is fit to the following functional form at different $J_{K}$, for all fillings:
 
 \begin{equation}
 E_{ED}(n)=E_{0}^{fit}(n)+(J_{K}/t)^{2}E_{2}^{fit}(n)+(J_{K}/t)^{4}E_{4}^{fit}(n)+.....
 \label{eq:get e2}
 \end{equation}
 
 with fit parameters $\{E_{0}^{fit}, E_{2}^{fit},E_{4}^{fit}\}$. Once we recover the set of $\{E_{2}^{fit}(n)\}$ for all spin configurations in the data base, we fit it to the following functional form
 
 \begin{equation}
 E_{2}^{fit}(n)=\varepsilon_{0}(n)+J_{1}^{fit}(n)\sum_{\langle ij \rangle}\mathbf{S}_{i} \cdot \mathbf{S}_{j}+J_{2}^{fit}(n) \sum_{\langle \langle ij \rangle \rangle} \mathbf{S}_{i} \cdot \mathbf{S}_{j}
 \label{eq:e2 fit}
 \end{equation}

by minimizing the norm of the following matrix equation

\begin{equation}
\underset{J_{1}^{fit}(n), J_{2}^{fit}(n),...}{\mbox{Min}} |\mathbf{M}(n) \cdot \mathbf{x}(n)-\mathbf{b}(n) |
\label{eq: matrix norm}
\end{equation}
 
 with $\{ J_{1}^{fit}(n), J_{2}^{fit}(n),...  \}$ as fit parameters. $\mathbf{M}$ is a $N_{s} \times (n_{J}+1)$, where $n_{J}$ is the number of couplings to be fit along with an additional constant $\varepsilon_{0}(n)$ in \eqref{eq:e2 fit}. The matrix $\mathbf{M}$ contains the classical energies corresponding to the couplings $J_{1,2,...}(n)$ for each random spin configuration, arranged along the rows. $\mathbf{x}$ is a vector of length $n_{J}+1$ given by $\mathbf{x}^{T}=(\varepsilon_{0}(n), J_{1}(n),J_{2}(n),...)$ and the vector $\mathbf{b}$ contains the extracted $E_{2}^{fit}(n)$ from \eqref{eq:get e2} for each of the $N_{s}$ spin configurations.
 
\begin{figure}[htpb]
\centering
\includegraphics[width=\linewidth]{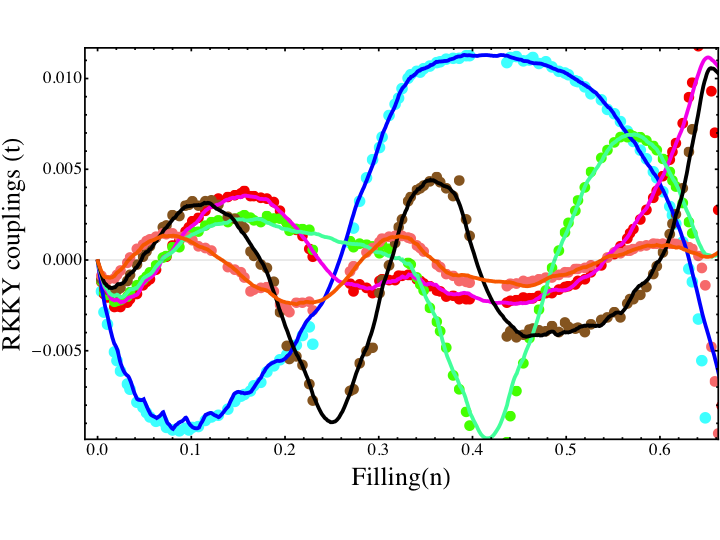} \caption{A comparison of RKKY interactions from 2 methods: (1) calculation by direct integration in Fourier space, shown via circles (2) Fitting procedure, shown via lines}
\label{fig5}
\end{figure}

A comparison of the Fourier and real space methods is shown for the first four RKKY couplings in Fig.\ref{fig5} as a function of filling $n$. Both methods have maximum susceptibility to finite size effects at small fillings $n \leq 0.1$ and to Ferromagnetic background orders where electrons have longer mean free paths comparable to system sizes. We next investigate, in more detail, the various spin orders making up the phase diagram in Fig.4 in \cite{maintext}.

\section{SUPPLEMENTARY INFORMATION S.I. II}

\begin{figure}[htpb]
\centering
\includegraphics[width=1\linewidth]{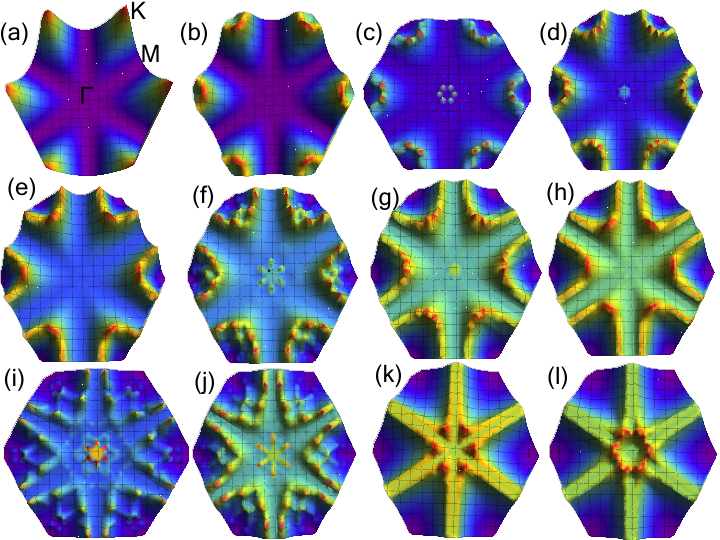} 
\caption{Eigenvalues $\lambda^{\nu}(\mathbf{q})$ of the $J_{\alpha \beta}(\mathbf{q})$ in the lowest band $\nu=1$(see Eq.3 in \cite{maintext}) in the Kagome first B.Z. (a) 0.333 (b)0.343 (c) 0.352 (d) 0.354 (e) 0.364 (f) 0.372 (g) 0.375 (h) 0.381 (i) 0.41. Data for $N=3\times 36^{2}$ lattice and color scheme is red(high), blue (low)}
\label{fig9}
\end{figure}

\begin{figure}[htpb]
\centering
\includegraphics[width=1\linewidth]{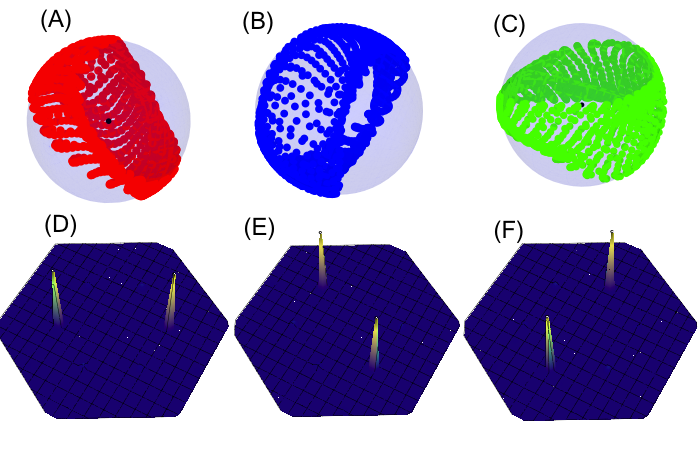} 
\caption{Common origin plot of spins and the Fourier weights for a spin configuration found at $n=0.375$ on a N = $3\times 36^{3}$ lattice. (A)-(C) common origin plots showing $(a_{1},b_{1},c_{1})$ phase. (D)-(F) Fourier profile $\eta_{\alpha}(\mathbf{q}$) (see Eq.4 in \cite{maintext}) for $\alpha =1,2,3$}
\label{fig10}
\end{figure}

The evolution of the Luttinger-Tisza matrix in the first B.Z. and the Fourier weights of a spin configuration from the highly symmetrical $(a_{1},b_{1},c_{1})$ phase is shown in Figs. \ref{fig9} and \ref{fig10} respectively. As the filling changes from $n=1/3$ to $n=0.41$, the ordering wave vector evolves smoothly from the zone corner to the zone center. At every filling, there are symmetry equivalent L.T. wave vectors which form the six fold star of Luttinger Tisza wave vectors. 

Correspondence between the star of symmetry related L.T. wave vectors and the ordering vectors of the corresponding spin configuration can be obtained by looking at the Fourier weights of the MC minimized spin pattern at every filling.  In our MC minimization, independent runs beginning from random spin configurations at 'high temperatures' relax to ground states which select different sets of wave vectors from the star of L.T. wave vectors. Fig. \ref{fig10} shows the Fourier weights in the first Brillouin zone for a spin pattern taken from a filling at $n=0.375$. The sharp peaks in the zone indicate that the order has a well defined wave vector content which can be detected experimentally in neutron scattering experiments.

\section{Supplementary Information S.I. III}
\label{sec:S2}
\begin{table}[ht]
\centering
\begin{tabular} {|c|c|c|c|c|}
 \hline\hline
Filling range & Phase Label  & non-coplanar? &  CO/ICO & Ex.\\ \hline
$(0,0.03)$  & $(a,a,a)$  & no & CO & FM\\ \hline  
$(0.115,0.146)$ & $(ab,bc,ca)$ & yes & ICO & Fig.\ref{fig7} \\ \hline
$\begin{array} {c} (0.226,0.25)\\ (0.492,0.551) \end{array}$  & $(a,b,c)$ & yes & ICO & Fig.\ref{fig11}  \\ \hline
$(0.325,0.333)$  & $(a_{1},a_{1},a_{1})$ & yes & ICO & Fig.3(a) in \cite{maintext} \\ \hline  
$\begin{array} {c} (0.362,0.485)\\ (0.579,0.624) \end{array}$  & $(a_{1},b_{1},c_{1})$ & yes & ICO & Fig.3(b) in \cite{maintext}  \\ \hline
  \hline 
 \end{tabular}
 \caption{Broken symmetry phases in RKKY limit. Columns from left to right: Filling range $n$ where the phase was found, Phase notation (see text), planar or non-coplanar orders, commensurate (CO) or incommensurate (ICO) order and example spin order}
\label{table:tab1}
 \end{table}

Spin orders with the same broken symmetries in Fourier space are classified as a single phase. The phase diagram in Fig.4 in \cite{maintext} shows eight such phases, each shown in a different color. The most prominent of these phases is also listed in Table \ref{table:tab1}. In this section, we take representative spin configurations from the most dominant phases in Fig.4 in \cite{maintext} and by looking at their Fourier space composition, illustrate why they form part of a smoothly connected second order phase. This section will also helps to establish the nomenclature for the different phases. 

We begin with the kind of phases where the dominant modes on each sublattice are made out of two of the six fold Luttinger-Tisza star of wave-vectors. The first kind of order has different weights of the modes on the three sublattices-- indicated as a $(ab,bc,ca)$ phase. The second phase is more symmetric as can be seen in Table \ref{table:tab2} and is labeled $(a_{1}b_{2},c_{1}a_{2},b_{1}c_{2})$ indicating two distinct $\eta_{\alpha}(\mathbf{q})$ magnitudes across all $\alpha$. A wave vector $a$ uses one of the two $\eta_{\alpha}$ values denoted by subscript (1) on  sublattice one and the second Fourier weight (2) on sublattice two. The same goes for the other wave-vectors $b,c$ highlighting the symmetry of the state.

\begin{table*}[ht]
\centering
\begin{tabular}{|p{2cm}|p{1.5cm}|p{1.5cm}|p{1.5cm}|p{1.5cm}|p{1.5cm}|p{1.5cm}|p{1.5 cm}|p{1.5 cm}|p{1.5cm}|}
\hline\hline
$Filling (n)$ & $\mathbf{q}_{1}(2\pi)$ & $\mathbf{q}_{2}(2\pi)$ & $\mathbf{q}_{3}(2\pi)$ & $\hspace{0.3cm} \eta_{1}(\mathbf{q}_{1})$ & $\hspace{0.3cm} \eta_{1}(\mathbf{q}_{2})$ & $\hspace{0.3cm} \eta_{2}(\mathbf{q}_{2})$ & $\hspace{0.3cm} \eta_{2}(\mathbf{q}_{3})$ & $\hspace{0.3cm} \eta_{3}(\mathbf{q}_{1})$ & $\hspace{0.3cm} \eta_{3}(\mathbf{q}_{3})$ \\
\hline
$\hspace{0.3cm}0.115$ & $\left (\begin{array}{c} 0.19 \\ -0.13 \end{array} \right)$ & $\left (\begin{array}{c} 0.02 \\ -0.22 \end{array} \right)$ & $\left (\begin{array}{c} 0.20 \\ 0.09 \end{array} \right)$ & $\hspace{0.3cm}0.22$ & $\hspace{0.3cm}0.22$ & $\hspace{0.3cm}0.28$ & $\hspace{0.3cm}0.19$ & $\hspace{0.3cm}0.29$ & $\hspace{0.3cm}0.19$\\
\hline
$\hspace{0.3cm}0.146$ & $\left (\begin{array}{c} 0.15 \\ -0.1 \end{array} \right)$ & $\left (\begin{array}{c} 0.02 \\ -0.18 \end{array} \right)$ & $\left (\begin{array}{c} 0.15 \\ -0.1 \end{array} \right)$ & $\hspace{0.3cm}0.21$ & $\hspace{0.3cm}0.21$ & $\hspace{0.3cm}0.27$ & $\hspace{0.3cm}0.19$ & $\hspace{0.3cm}0.28$ & $\hspace{0.3cm}0.19$\\
\hline
$\hspace{0.3cm}0.181$ & $\left (\begin{array}{c} 0.02 \\ -0.13 \end{array} \right)$ & $\left (\begin{array}{c} 0.1 \\ 0.08 \end{array} \right)$ & $\left (\begin{array}{c} -0.13 \\ 0.05 \end{array} \right)$ & $\hspace{0.3cm}0.22$ & $\hspace{0.3cm}0.22$ & $\hspace{0.3cm}0.22$ & $\hspace{0.3cm}0.22$ & $\hspace{0.3cm}0.22$ & $\hspace{0.3cm}0.22$\\
\hline
\end{tabular}
\caption{Dominant wave vectors and their Fourier weights on the sublattices. Left to right: Filling at which the spin order originates in the RKKY limit, dominant wave vectors in the first B.Z., weights of the dominant wave vectors on each of the three sublattices. Fillings $0.115$ and $0.146$ correspond to $(ab,bc,ca)$ phase, while the more symmetric spin order at $0.181$ is a spin set from the $(a_{1}b_{2},c_{1}a_{2},b_{1}c_{2})$ phase. Spin configurations are shown in Fig. \ref{fig7}}
\label{table:tab2}
\end{table*}

The next kind of phase is where each sublattice is made out of its own independent single dominant Luttinger-Tisza wave vector. There are again two types of such phases shown in Fig.\ref{fig11}. One in which the Fourier weights of each of the three wave-vectors $a,b,c$ is different on each sublattice. This phase is labeled as $(a,b,c)$. The second type of phase is a highly symmetric version of the former type. Each sublattice has the same weight of the dominant Fourier mode and is labeled $(a_{1},b_{1},c_{1})$. The Fourier components and weights of the spin orders from these two types of phases are enlisted in Table \ref{table:tab4}

The next category of phases are the $1\mathbf{Q}$ type orders, where all sublattices are made out of the same dominant wave-vector. These again fall in to a symmetric --$(a_{1},a_{1},a_{1})$ and an asymmetric version $(a,a,a)$. Fourier weights of the dominant mode of spin orders from these two types of phases is shown in Table \ref{table:tab3}

\begin{table}[ht]
\centering
\begin{tabular}{|p{2cm}|p{1.5cm}|p{1.5cm}|p{1.5cm}|p{1.5cm}|}
\hline\hline
$Filling (n)$ & $\hspace{0.3cm}\mathbf{q}(2\pi)$ & $\hspace{0.4cm}\eta_{1}$ & $\hspace{0.4cm}\eta_{2}$ & $\hspace{0.4cm}\eta_{3}$ \\
\hline
$0.321$ &  $\left (\begin{array}{c} 0.27 \\ 0.04 \end{array} \right)$ & $\hspace{0.4cm}0.47$ & $\hspace{0.4cm}0.47$ & $\hspace{0.4cm}0.45$ \\
\hline
$0.325$ &  $\left (\begin{array}{c} 0.14 \\ 0.29 \end{array} \right)$ & $\hspace{0.4cm}0.497$ & $\hspace{0.4cm}0.497$ & $\hspace{0.4cm}0.497$ \\
\hline
$1/3$ &  $\hspace{0.4cm}\mathbf{K}$ & $\hspace{0.4cm}1$ & $\hspace{0.4cm}1$ & $\hspace{0.4cm}1$ \\
\hline

\end{tabular}
\caption{Dominant wave vectors and their Fourier weights on the sublattices. Left to right: Filling at which the spin order originates in the RKKY limit, dominant wave vector in the first B.Z., weight of the dominant wave vectors on each of the three sublattices. Spin order at filling $0.321$ corresponds to $(a,a,a)$ phase. The more symmetric spin orders found at $0.325$ (coplanar spiral Fig.3 in \cite{maintext}) and $1/3$ ($\sqrt{3}\times \sqrt{3}$ order Fig.1 in \cite{maintext}) form part of the $(a_{1},a_{1},a_{1})$ phase}
\label{table:tab3}
\end{table}

\begin{table*}[ht]
\centering
\begin{tabular}{|p{0.8cm}|p{1.41cm}|p{1.41cm}|p{1.41cm}|p{2.3cm}|p{2.3cm}|p{2.3cm}|p{1cm}|p{1cm}|p{1cm}|}
\hline\hline
$n$ & $\hspace{3mm}\mathbf{q}_{1}(2\pi)$ & $\hspace{3mm}\mathbf{q}_{2}(2\pi)$ & $\hspace{3mm}\mathbf{q}_{3}(2\pi)$ & $\hspace{5mm}\mathbf{S}_{1}(\mathbf{q}_{1})$ & $\hspace{5mm}\mathbf{S}_{2}(\mathbf{q}_{2})$ & $\hspace{5mm}\mathbf{S}_{3}(\mathbf{q}_{3})$ & $\eta_{1}(\mathbf{q}_{1})$ &$\eta_{2}(\mathbf{q}_{2})$ & $\eta_{3}(\mathbf{q}_{3})$\\
\hline
$0.226$ & $\left (\begin{array}{c} 0.25 \\ 0.12 \end{array} \right)$ & $\left (\begin{array}{c} -0.25 \\ 0.12 \end{array} \right)$ & $\left (\begin{array}{c} 0.02 \\ -0.27 \end{array} \right)$ & $\left (\begin{array}{c} 0.29e^{i 1.38} \\ 0.38 e^{i 1.48 } \\ 0.48 e^{-i 0.18} \end{array}\right)$ & $\left (\begin{array}{c} 0.26e^{-i 1.82} \\ 0.38 e^{-i 1.61 } \\ 0.46 e^{i 2.95} \end{array}\right)$ & $\left (\begin{array}{c} 0.45e^{i 1.48} \\ 0.39 e^{-i 0.48 } \\ 0.30 e^{i 0.29} \end{array}\right)$ & $0.92$ & $0.86$ & $0.90$ \\
\hline

$0.341$ & $\left (\begin{array}{c} 0.12 \\ -0.24 \end{array} \right)$ & $\left (\begin{array}{c} 0.12 \\ 0.24 \end{array} \right)$ & $\left (\begin{array}{c} -0.27 \\ 0.01 \end{array} \right)$ & $\left (\begin{array}{c} 0.2e^{-i 2.1} \\ 0.41 e^{-i 1.36 } \\ 0.2 e^{i 1.87} \end{array}\right)$ & $\left (\begin{array}{c} 0.49e^{i 1.54} \\ 0.18 e^{-i 2.5 } \\ 0.14 e^{i 0.97} \end{array}\right)$ & $\left (\begin{array}{c} 0.06e^{-i 1.7} \\ 0.24 e^{i 1.44 } \\ 0.47 e^{i 1.44} \end{array}\right)$ & $0.5$ & $0.58$ & $0.57$ \\
\hline
$0.355$ & $\left (\begin{array}{c} 0.1 \\ -0.2 \end{array} \right)$ & $\left (\begin{array}{c} 0.1 \\ 0.2 \end{array} \right)$ & $\left (\begin{array}{c} -0.23 \\ 0.01 \end{array} \right)$ & $\left (\begin{array}{c} 0.18e^{i 2.04} \\ 0.46 e^{i 1.76 } \\ 0.22 e^{-i 2.28} \end{array}\right)$ & $\left (\begin{array}{c} 0.42e^{-i 0.21 } \\ 0.19 e^{-i 2.6 } \\ 0.42 e^{-i 1.7} \end{array}\right)$ & $\left (\begin{array}{c} 0.43e^{-i 0.48} \\ 0.21 e^{-i 2.88 } \\ 0.44 e^{-i 1.94} \end{array}\right)$ & $0.6$ & $0.8$ & $0.84$ \\
\hline
$0.492$ & $\left (\begin{array}{c} 0.1 \\ 0.13 \end{array} \right)$ & $\left (\begin{array}{c} 0.1 \\ -0.13 \end{array} \right)$ & $\left (\begin{array}{c} 0.06 \\ 0.16 \end{array} \right)$ & $\left (\begin{array}{c} 0.27e^{i 0.94} \\ 0.43 e^{i 1.48 } \\ 0.47 e^{-i 0.26} \end{array}\right)$ & $\left (\begin{array}{c} 0.45e^{-i 3.02 } \\ 0.28 e^{i } \\ 0.43 e^{i 1.84} \end{array}\right)$ & $\left (\begin{array}{c} 0.27e^{-i 2.05} \\ 0.41 e^{-i 2.55 } \\ 0.46 e^{-i 0.81} \end{array}\right)$ & $0.96$ & $0.92$ & $0.92$ \\
\hline
$0.527$ & $\left (\begin{array}{c} 0.23 \\ 0.16 \end{array} \right)$ & $\left (\begin{array}{c} 0.23 \\ -0.15 \end{array} \right)$ & $\left (\begin{array}{c} 0.02 \\ 0.27 \end{array} \right)$ & $\left (\begin{array}{c} 0.47e^{i 1.98} \\ 0.48 e^{i 0.3 } \\ 0.2 e^{i 1.23} \end{array}\right)$ & $\left (\begin{array}{c} 0.47e^{-i 1.87 } \\ 0.17 e^{-i 2.55 } \\ 0.49 e^{-i 0.34} \end{array}\right)$ & $\left (\begin{array}{c} 0.46e^{-i 1.98} \\ 0.47 e^{-i 0.3 } \\ 0.21 e^{-i 1.19} \end{array}\right)$ & $0.98$ & $0.98$ & $0.96$ \\
\hline
$0.551$ & $\left (\begin{array}{c} 0.21 \\ 0.1 \end{array} \right)$ & $\left (\begin{array}{c} -0.2 \\ 0.1 \end{array} \right)$ & $\left (\begin{array}{c} 0.02 \\ -0.22 \end{array} \right)$ & $\left (\begin{array}{c} 0.01e^{-i 2.17} \\ 0.49 e^{i 1.15 } \\ 0.47 e^{i 2.7} \end{array}\right)$ & $\left (\begin{array}{c} 0.45e^{-i 2.86 } \\ 0.49 e^{-i 1.35 } \\ 0.19 e^{i 3.10} \end{array}\right)$ & $\left (\begin{array}{c} 0.48e^{i 0.39} \\ 0.03 e^{-i 2.91 } \\ 0.47 e^{i 1.98} \end{array}\right)$ & $0.92$ & $0.94$ & $0.9$ \\
\hline\hline

$0.28$ & $\left (\begin{array}{c} 0.125 \\ 0.12 \end{array} \right)$ & $\left (\begin{array}{c} -0.17 \\ 0.04 \end{array} \right)$ & $\left (\begin{array}{c} 0.04 \\ -0.17 \end{array} \right)$ & $\left (\begin{array}{c} 0.25e^{-i 1.69} \\ 0.49 e^{i 3.02 } \\ 0.43 e^{-i 1.69} \end{array}\right)$ & $\left (\begin{array}{c} 0.41e^{i 0.43} \\ 0.44 e^{-i 1.47 } \\ 0.35 e^{i 2.68} \end{array}\right)$ & $\left (\begin{array}{c} 0.48e^{i 1.2} \\ 0.28 e^{i 2.5 } \\ 0.41 e^{i 2.9} \end{array}\right)$ & $0.96$ & $0.96$ & $0.96$ \\

\hline
$0.362$ & $\left (\begin{array}{c} 0.1 \\ -0.15 \end{array} \right)$ & $\left (\begin{array}{c} 0.08 \\ 0.16 \end{array} \right)$ & $\left (\begin{array}{c} 0.18 \\ 0.01 \end{array} \right)$ & $\left (\begin{array}{c} 0.43e^{i 0.16} \\ 0.4 e^{-i 1.51 } \\ 0.22 e^{-i 1.06} \end{array}\right)$ & $\left (\begin{array}{c} 0.36e^{-i 2.31 } \\ 0.29 e^{-i 1.76 } \\ 0.42 e^{i 2.6} \end{array}\right)$ & $\left (\begin{array}{c} 0.27e^{-i 2.48} \\ 0.4 e^{i 1.55 } \\ 0.4 e^{-i 0.27} \end{array}\right)$ & $0.8$ & $0.8$ & $0.8$ \\
\hline
$0.394$ & $\left (\begin{array}{c} 0.06 \\ -0.08 \end{array} \right)$ & $\left (\begin{array}{c} 0.04 \\ 0.09 \end{array} \right)$ & $\left (\begin{array}{c} 0.1 \\ 0.01 \end{array} \right)$ & $\left (\begin{array}{c} 0.47e^{i 2.78} \\ 0.1 e^{i 2.34 } \\ 0.47 e^{i 1.2} \end{array}\right)$ & $\left (\begin{array}{c} 0.38e^{i 0.74 } \\ 0.47 e^{i 2.44 } \\ 0.29 e^{i 1.1} \end{array}\right)$ & $\left (\begin{array}{c} 0.36e^{-i 1.01} \\ 0.44 e^{i 0.85 } \\ 0.34 e^{i 2.75} \end{array}\right)$ & $0.9$ & $0.9$ & $0.9$ \\
\hline
$0.485$ & $\left (\begin{array}{c} 0.1 \\ 0.08 \end{array} \right)$ & $\left (\begin{array}{c} -0.12 \\ 0.04 \end{array} \right)$ & $\left (\begin{array}{c} 0.02 \\ -0.13 \end{array} \right)$ & $\left (\begin{array}{c} 0.44e^{i 1.71} \\ 0.46 e^{i 0.24 } \\ 0.22 e^{i 2.38} \end{array}\right)$ & $\left (\begin{array}{c} 0.33e^{i 1.45 } \\ 0.44 e^{i 2.54 } \\ 0.4 e^{-i 2.5} \end{array}\right)$ & $\left (\begin{array}{c} 0.45e^{-i 1.5} \\ 0.17 e^{i 1.6 } \\ 0.48 e^{-i 3.06} \end{array}\right)$ & $0.92$ & $0.92$ & $0.92$ \\
\hline
$0.579$ & $\left (\begin{array}{c} 0.25 \\ 0.1 \end{array} \right)$ & $\left (\begin{array}{c} 0.22 \\ -0.15 \end{array} \right)$ & $\left (\begin{array}{c} 0.02 \\ 0.27 \end{array} \right)$ & $\left (\begin{array}{c} 0.44e^{-i 2.12} \\ 0.28 e^{i 1.76 } \\ 0.47 e^{-i 0.36} \end{array}\right)$ & $\left (\begin{array}{c} 0.38e^{-i 1.88 } \\ 0.41 e^{-i 0.91 } \\ 0.42 e^{i 0.21} \end{array}\right)$ & $\left (\begin{array}{c} 0.39e^{-i 0.62} \\ 0.49 e^{i 0.94 } \\ 0.31 e^{-i 0.65} \end{array}\right)$ & $0.98$ & $0.98$ & $0.98$ \\
\hline
$0.596$ & $\left (\begin{array}{c} 0.23 \\ 0.15 \end{array} \right)$ & $\left (\begin{array}{c} 0.25 \\ -0.12 \end{array} \right)$ & $\left (\begin{array}{c} 0.02 \\ 0.27 \end{array} \right)$ & $\left (\begin{array}{c} 0.48e^{-i 2.78} \\ 0.19 e^{i 2.6 } \\ 0.47 e^{i 1.85} \end{array}\right)$ & $\left (\begin{array}{c} 0.48e^{-i 2.53 } \\ 0.47 e^{i 2.08 } \\ 0.21 e^{i -0.44} \end{array}\right)$ & $\left (\begin{array}{c} 0.48e^{i 2.77} \\ 0.2 e^{-i 2.55 } \\ 0.47 e^{-i 1.85} \end{array}\right)$ & $0.98$ & $0.98$ & $0.98$ \\
\hline
$0.624$ & $\left (\begin{array}{c} 0.17 \\ 0.12 \end{array} \right)$ & $\left (\begin{array}{c} -0.18 \\ 0.08 \end{array} \right)$ & $\left (\begin{array}{c} 0.02 \\ -0.2 \end{array} \right)$ & $\left (\begin{array}{c} 0.48e^{-i 1.46} \\ 0.32 e^{-i 0.1 } \\ 0.38 e^{-i 2.87} \end{array}\right)$ & $\left (\begin{array}{c} 0.35 e^{-i 0.31 } \\ 0.37 e^{i 2.2 } \\ 0.47 e^{-i 2.21} \end{array}\right)$ & $\left (\begin{array}{c} 0.17e^{i 2.2} \\ 0.49 e^{-i 2.8 } \\ 0.46 e^{i 1.86} \end{array}\right)$ & $0.96$ & $0.96$ & $0.9$ \\ \hline
\end{tabular}
\caption{Dominant wave-vectors and their Fourier weights on the sublattices for representative spin orders from the phases --$(a,b,c)$ and $(a_{1},b_{1},c_{1})$. Corresponding spin configurations are shown in Fig. \ref{fig11}}
\label{table:tab4}
\end{table*}

\section{Supplementary Information S.I. IV}
\label{sec:S3}

A parametrization of the different spin orders $\{ \mathbf{S}^{opt}_{i}\}$ making up the phases in Fig.4 in \cite{maintext} is provided in this section. There are two steps for parametrizing a spin state. In the first step, we construct a purified version $\{ \mathbf{S}^{recons}_{i(\alpha)}\}$ of the optimal spin order obtained from MC by filtering out the dominant Fourier modes on each sublattice. In the second step, we try fitting simple functional forms to $\{ \mathbf{S}^{recons}_{i(\alpha)}\}$. Since, most of the states are combinations of simple coplanar incommensurate spirals, we begin by parametrizing coplanar spirals. 

A coplanar spiral is parametrized by its ordering wave vector $\mathbf{q}$ and two phases dependent on the locking between the sublattices. For two orthonormal unit vectors $\mathbf{e}_{1,2}$, a coplanar spiral can be parametrized as 

\begin{equation}
\mathbf{S}_{i(\alpha)}=\mathbf{Re}[e^{i (\mathbf{q} \cdot \mathbf{R}_{i}+\varphi_{\alpha})}(\mathbf{e}_{1}-i \mathbf{e}_{2} )]
\label{eq:cs}
\end{equation}

Depending on the location of $\mathbf{q}$ in the zone, the spiral can further be classified as commensurate - if $\mathbf{q}$ lies at a special symmetry point in the zone) or incommensurate - if $\mathbf{q}$ lies at an arbitrary wave vector. Examples of special commensurate coplanar spirals are the two well known $\mathbf{q}=0$ and the $\sqrt{3}\times \sqrt{3}$ order with ordering wave vectors $\mathbf{q}=\mathbf{\Gamma}$ and $\mathbf{q}=\mathbf{K}$ lying at the zone center and zone corner, respectively. For both orders, $\varphi_{\alpha}=2\pi (\alpha-1)/3$ which makes an angle of $2\pi/3$ between spins, locally, on every triangle. 

An incommensurate order, on the other hand, has a special direction in real space defined by $\mathbf{q}$ and a set of points $\{\mathbf{R}_{i(\alpha)} \}$, such that $\mathbf{q} \cdot \mathbf{R}_{i(\alpha)}=2\pi m(\sqrt{N/3})^{-1}$ for an integer $m$. As we move along $\{\mathbf{R}_{i(\alpha)} \}$, we trace out $(\sqrt{N/3})^{-1}$ equally spaced coplanar directions in spin space as shown in Fig.3(a1)-(a3) in \cite{maintext}. An incommensurate coplanar spiral might locally have angles close to $2\pi/3$ as in the twisted $\sqrt{3}\times \sqrt{3}$ order -Fig.3 in \cite{maintext} and Table \ref{table:tab3} . As discussed in the previous section, the two kinds of spiral orders can belong to an $(a,a,a)$ phase or a more symmetric $(a_{1},a_{1},a_{1})$ phase. States from these phases are shown in Fig.\ref{fig6} below. 

\begin{figure}[htpb]
\centering
\includegraphics[width=\linewidth]{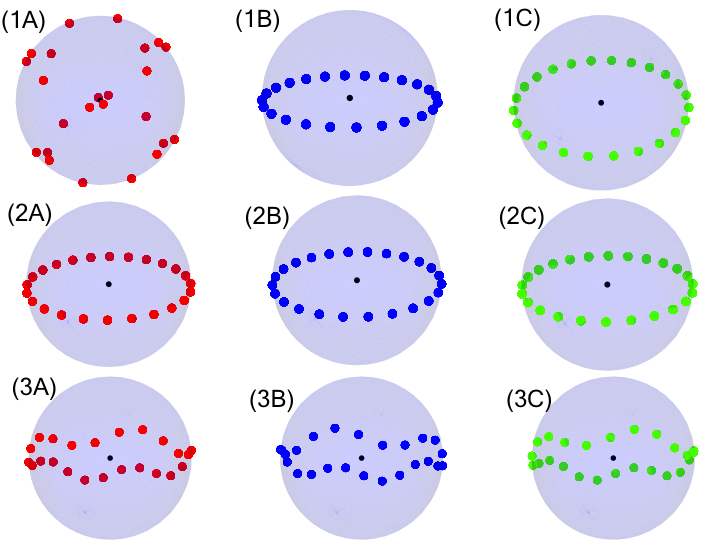} 
\caption{Common origin plot of incommensurate coplanar spiral orders belonging to $(a,a,a)$ and $(a_{1},a_{1},a_{1})$ phases in Fig. 4 in \cite{maintext}. 1(A)-(C): spins on each of the three sublattices for a spin order recovered from MC at $n=0.311$,2(A)-(C): at $n=0.321$,3(A)-(C): at $n=0.325$. The ordering wave vectors for the three states in Table \ref{table:tab3} and Table \ref{table:tab5} along with Eq.\ref{eq:cs} can be used to construct $\{ \mathbf{S}^{recons}_{i(\alpha)} \}$ }
\label{fig6}
\end{figure}

We now consider the more complicated $3\mathbf{Q}$ (ab,bc,ca) orders in the phase diagram in Fig.4 \cite{maintext}. The simplest of these orders are the two commensurate \emph{cuboc1}\cite{SI:messio1} and \emph{cuboc2}\cite{SI:domenge} orders shown in Fig.1 in \cite{maintext} and found at fillings $n=5/12$ and $n=2/3$. Each of these orders are made from the three ordering vectors $\mathbf{Q}_{1,2,3} \in \mathbf{M}$ belonging to the zone mid points and leading to a twelve site magnetic unit cell. Each sublattice $\alpha$ uses only two of the three vectors leading to the label $3\mathbf{Q}$ (ab,bc,ca). \emph{Cuboc1} is parametrized as:



\begin{figure}[htpb]
\centering
\includegraphics[width=\linewidth]{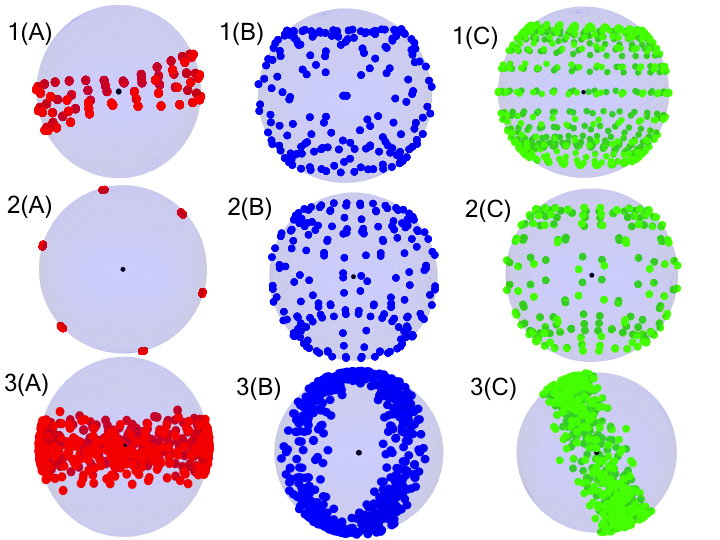} 
\caption{Common origin plot of spin orders from the $(ab,bc,ca)$ and $(a_{1}b_{2},c_{1}a_{2},b_{1}c_{2})$ phases in Fig. 4 in \cite{maintext}. 1(A)-(C): spins on each of the three sublattices for a spin order recovered from MC at $n=0.115$,2(A)-(C): at $n=0.146$,3(A)-(C): at $n=0.181$. The ordering wave vectors for the three states in Table \ref{table:tab7} and $\mathbf{S}_{\alpha}(\mathbf{q})$ along with \ref{eq:s3q} can be used to construct $\{ \mathbf{S}^{recons}_{i} \}$ }.
\label{fig7}
\end{figure}

\begin{align}
\mathbf{S}_{i(1)} &=\frac{1}{\sqrt{2}} [\cos(\mathbf{Q}_{2} \cdot \mathbf{R}_{i}) \mathbf{e}_{2}+\cos(\mathbf{Q}_{3} \cdot \mathbf{R}_{i}) \mathbf{e}_{3}   ] \nonumber \\
\mathbf{S}_{i(2)} &=\frac{1}{\sqrt{2}} [\cos(\mathbf{Q}_{1} \cdot \mathbf{R}_{i}) \mathbf{e}_{1}-\cos(\mathbf{Q}_{3} \cdot \mathbf{R}_{i}) \mathbf{e}_{3}   ]\nonumber  \\
\mathbf{S}_{i(3)} &=-\frac{1}{\sqrt{2}} [\cos(\mathbf{Q}_{1} \cdot \mathbf{R}_{i}) \mathbf{e}_{1}+\cos(\mathbf{Q}_{2} \cdot \mathbf{R}_{i}) \mathbf{e}_{2} ]\nonumber \\
\label{eq:cuboc1}
\end{align}
where $\mathbf{Q}_{2}=2\pi(1/4,-1/(4\sqrt{3}))$, $\mathbf{Q}_{3}=\mathcal{R}_{\pi/3}\mathbf{Q}_{2}$ and $\mathbf{Q}_{1}=\mathcal{R}_{2\pi/3}\mathbf{Q}_{2}$. $\mathcal{R}_{\theta}$ is the $2 \times 2$ rotation matrix. The state does not elicit an Anomalous Hall response due to the coplanarity of spins on every triangle.

Nearest neighbor spins in the \emph{Cuboc2} state make an angle of $\pi/3$, while the next nearest neighbor spins have an angle of $2\pi/3$ between them. The state is thus favored in the presence of a ferromagnetic $J_{1}$ and an AFM $J_{2}$ interaction\cite{SI:domenge}. The non-coplanarity of spins within each triangle in \emph{cuboc2} leads to a non-vanishing value of the scalar spin chirality $\chi=\mathbf{S}_{i} \cdot (\mathbf{S}_{j}\times \mathbf{S}_{k})$. $\chi$ is $+(-)1/\sqrt{2}$ on all the up (down) triangles. The equal and opposite fluxes leads to zero overall flux and no Anomalous Hall response.The state has a $Z_{2}$ symmetric partner and at zero temperature, one of the two states is spontaneously selected, breaking $Z_{2}$ symmetry. The spin order is parametrized as follows:

\begin{align}
\mathbf{S}_{i(1)} &=\frac{1}{\sqrt{2}} [\cos(\mathbf{Q}_{2} \cdot \mathbf{R}_{i}) \mathbf{e}_{2}+\cos(\mathbf{Q}_{3} \cdot \mathbf{R}_{i}) \mathbf{e}_{3}   ] \nonumber \\
\mathbf{S}_{i(2)} &=\frac{1}{\sqrt{2}} [-\cos(\mathbf{Q}_{1} \cdot \mathbf{R}_{i}) \mathbf{e}_{1}+\cos(\mathbf{Q}_{3} \cdot \mathbf{R}_{i}) \mathbf{e}_{3}   ]\nonumber  \\
\mathbf{S}_{i(3)} &=\frac{1}{\sqrt{2}} [-\cos(\mathbf{Q}_{1} \cdot \mathbf{R}_{i}) \mathbf{e}_{1}+\cos(\mathbf{Q}_{2} \cdot \mathbf{R}_{i}) \mathbf{e}_{2} ] \nonumber \\
\end{align}

Parametrization of other constituent states of the $(ab,bc,ca)$ type phase, such as the spin configurations shown in Fig \ref{fig7} is done using the Fourier transform of the spin orders $\{  \mathbf{S}_{\alpha}(\mathbf{q}_{\mu}) \}$ from MC minimization. The "purified" order is obtained by simply inverse F.T. the vectors $\{  \mathbf{S}_{\alpha}(\mathbf{q}_{\mu}) \}$, given in Table \ref{table:tab7}, using Eq. \ref{eq:s3q}. \begin{equation}
\mathbf{S}^{recons}_{i(\alpha)}=\mathcal{N}_{i} \sum_{\mathbf{q} \in \{\mathbf{q}_{\mu} \}}\mathbf{S}_{\alpha}e^{i \mathbf{q} \cdot \mathbf{R}_{i}}
\label{eq:s3q}
\end{equation}

The ordering wave vectors and spins in Fourier space for the three orders in Fig.\ref{fig7} is given in Table \ref{table:tab7}.

We next turn to spin orders from the  $(a,b,c)$ phase. For \emph{all} orders in this phase, spins on the three sublattices are defined by mutually exclusive wave vectors tracing out coplanar spirals. For \emph{most} spin orders, the three sublattice dependent planes, are mutually orthogonal. Spins are parametrized as: 

\begin{align}
\mathbf{S}^{recons}_{i(\alpha=1)} &=\mathbf{Re}[e^{(\mathbf{q}_{1}.\mathbf{R}_{i}+\varphi_{1})}(\mathbf{e}_{1}-i\mathbf{e}_{2})] \nonumber \\
\mathbf{S}^{recons}_{i(\alpha=2)} &=\mathbf{Re}[e^{(\mathbf{q}_{2}.\mathbf{R}_{i}+\varphi_{2})}(\mathbf{e}_{2}-i\mathbf{e}_{3})] \nonumber \\
\mathbf{S}^{recons}_{i(\alpha=3)} &=\mathbf{Re}[e^{\mathbf{q}_{3}.\mathbf{R}_{i}+\varphi_{3})}(\mathbf{e}_{1}-i\mathbf{e}_{3})] \nonumber \\
 \label{eq:3Q}
\end{align}

where $\mathbf{e}_{1,2,3}$ form a triad of orthonormal vectors (see Table \ref{table:tab8} ). For a few spin orders at $n=0.226,0.228$ and at $n=0.527$ from within this phase, spins on two of the three sublattices lie in the same plane perpendicular to the plane in which spins on the third sublattice lie.

\begin{figure}[htpb]
\centering
\includegraphics[width=1\linewidth]{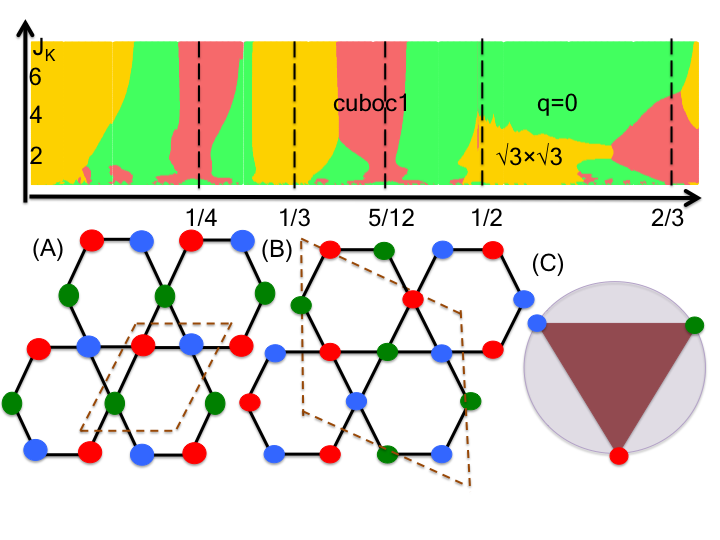} 
\caption{Variational phase diagram in the presence of a nearest neighbor antiferromagnetic exchange interaction $J_{ex}=10t$. (A): $\mathbf{q}=0$, (B): $\sqrt{3}\times \sqrt{3}$ (C): Common origin plot of spins for orders in (A),(B)}
\label{fig8}
\end{figure}

\begin{table}[ht]
\centering
\begin{tabular} {|p{1.5cm}|p{2.2cm}|p{2cm}|p{2.2cm}|}
\hline\hline
$\hspace{4mm}\mathbf{q}(2\pi)$ & $\hspace{4mm} \mathbf{S}_{1}(\mathbf{q})$ & $\hspace{4mm} \mathbf{S}_{2}(\mathbf{q})$ & $\hspace{4mm} \mathbf{S}_{3}(\mathbf{q})$ \\
\hline\hline
 $\left (\begin{array}{c} 0.12 \\ -0.19 \end{array} \right)$ & $\left (\begin{array}{c} 0.32e^{-i 2.9} \\ 0.4e^{i 1.55} \\ 0.26e^{i 2.8} \end{array} \right)$ &  $\left (\begin{array}{c} 0.37e^{i 0.9} \\ 0.45e^{-i 0.8} \\ 0.3e^{i 0.4} \end{array} \right)$ &  $\left (\begin{array}{c} 0.36e^{-i 0.5} \\ 0.43e^{-i 2.2} \\ 0.29e^{-i} \end{array} \right)$ \\
\hline\hline
\end{tabular}
\caption{Ordering wave vectors and spin F.T. for reconstructing order at $n=0.311$ occurring in the phase diagram in Fig. 4 in \cite{maintext}. From left to right: order number corresponding to labeling in Fig.4 in \cite{maintext}, Ordering wave vector $\mathbf{q}$ for the $1\mathbf{Q}(a,a,a)$ state, spin F.T. at each wave vector.The spin order Fig.\ref{fig6} has an additional significant contribution ($\sim 30\%$) on sublattice $\alpha=1$ from an additional wave vector $\mathbf{q}_{2}=2\pi(0.12,0.19)$ with Fourier weight $\mathbf{S}_{\alpha=1}(\mathbf{q}_{2})=(0.25e^{i1.93},0.1 e^{i1.96},0.3e^{-i 1.19})$}
\label{table:tab5}
\end{table}

\begin{table}[ht]
\centering
\begin{tabular}{|p{2cm}|p{1.5cm}|p{1cm}|p{1cm}|p{2cm}|}
\hline\hline
$Filling (n)$ & $\hspace{4mm}\mathbf{q}(2\pi)$ & $\varphi_{2} (2\pi)$ & $ \varphi_{3} (2\pi)$ & $\mbox{Norm} \{\mathbf{S}^{recons}_{i}\}$ \\
\hline
$0.321$ & $\left (\begin{array}{c} 0.27 \\ 0.04 \end{array} \right)$ & $0.27$ & $2/3$ & (0.95,0.97)\\
\hline
$0.325$ & $\left (\begin{array}{c} 0.14 \\ 0.29 \end{array} \right)$ & $0.65$ & $0.27$ & (0.997,0.998) \\
\hline\hline
\end{tabular}
\caption{Ordering wave vectors and spin F.T. for reconstructing orders found at $n=0.321$ and $n=0.325$. From left to right: filling at which order was found, ordering wave vector $\mathbf{q}$ for the $1\mathbf{Q}(a,a,a)$ state, sublattice phases $\varphi_{2,3}$ (see \eqref{eq:cs}) and $\mbox{Norm} \{\mathbf{S}^{recons}_{i}\}$}
\label{table:tab6}
\end{table}

\section{Supplementary Information S.I. V}
\label{sec:S4}

Here we show that the \emph{Cuboc1} state is also selected from the family of degenerate states of the nearest neighbor HAF on Kagome by turning on the KLM Hamiltonian. To this end, we add a strong nearest neighbor antiferromagnetic interaction of strength $J_{ex}$ to the KLM Hamiltonian and using exact diagonalization explore the stability of states within the variational approach outlined before. 

Fig.\ref{fig8} shows the phase diagram in the presence of a strong $J_{ex}=10t$ which suppresses all the incommensurate orders and selects within the manifold of the three $120$ degree states- the $\mathbf{q}=0$, $\sqrt{3} \times \sqrt{3}$ and the \emph{cuboc1} order. As can be seen, large parts of the phase diagram are dominated by the \emph{cuboc1} state.

\begin{table}[ht]
\centering
\begin{tabular}{|p{0.5cm}|p{1.39cm}|p{1.39cm}|p{1.39cm}|p{1cm}|p{1cm}|p{1cm}|}
\hline\hline
$\#$ & $\hspace{4mm}\mathbf{q}_{1}(2\pi)$ & $\hspace{4mm}\mathbf{q}_{2}(2\pi)$ & $\hspace{4mm}\mathbf{q}_{3}(2\pi)$ & $\varphi_{1} (2\pi)$ & $\varphi_{2} (2\pi)$ & $ \varphi_{3} (2(\pi)$ \\
\hline
$(6)$ & $\left (\begin{array}{c} 0.125 \\ 0.12 \end{array} \right)$ &$\left (\begin{array}{c} -0.17 \\ 0.05 \end{array} \right)$ &$\left (\begin{array}{c} 0.05 \\ -0.17 \end{array} \right)$  & $-0.91$ & $-1.36$ & $-0.86$\\
\hline\hline
$(12)$ & $\left (\begin{array}{c} 0.1 \\ -0.15 \end{array} \right)$ &$\left (\begin{array}{c} 0.08 \\ 0.17 \end{array} \right)$ &$\left (\begin{array}{c} 0.19 \\ 0.01 \end{array} \right)$  & $-2.04$ & $-0.58$ & $0.99$\\
\hline\hline
$(13)$ & $\left (\begin{array}{c} 0.06 \\ -0.08 \end{array} \right)$ &$\left (\begin{array}{c} -0.04 \\ -0.09 \end{array} \right)$ &$\left (\begin{array}{c} 0.1 \\ 0.01 \end{array} \right)$  & $0.39$ & $1.9$ & $1$\\
\hline\hline
$(14)$ & $\left (\begin{array}{c} 0.1 \\ 0.08 \end{array} \right)$ &$\left (\begin{array}{c} -0.12 \\ -0.05 \end{array} \right)$ &$\left (\begin{array}{c} 0.02 \\ -0.13 \end{array} \right)$  & $0.91$ & $2.26$ & $1.1$\\
\hline\hline
$(16)$ & $\left (\begin{array}{c} 0.22 \\ 0.15 \end{array} \right)$ &$\left (\begin{array}{c} 0.22 \\ -0.15 \end{array} \right)$ &$\left (\begin{array}{c} 0.02 \\ 0.27 \end{array} \right)$  & $-0.12$ & $-1.53$ & $-3.02$\\
\hline\hline
$(17)$ & $\left (\begin{array}{c} 0.21 \\ 0.1 \end{array} \right)$ &$\left (\begin{array}{c} -0.2 \\ -0.1 \end{array} \right)$ &$\left (\begin{array}{c} 0.02 \\ -0.23 \end{array} \right)$  & $-0.48$ & $1.79$ & $-0.97$\\
\hline\hline
$(18)$ & $\left (\begin{array}{c} 0.25 \\ 0.12 \end{array} \right)$ &$\left (\begin{array}{c} 0.23 \\ -0.16 \end{array} \right)$ &$\left (\begin{array}{c} -0.02 \\ -0.28 \end{array} \right)$  & $1.8$ & $0.64$ & $-0.37$\\
\hline\hline
$(19)$ & $\left (\begin{array}{c} 0.22 \\ 0.15 \end{array} \right)$ &$\left (\begin{array}{c} -0.25 \\ 0.12 \end{array} \right)$ &$\left (\begin{array}{c} 0.02 \\ 0.27 \end{array} \right)$  & $0.97$ & $-0.35$ & $0.96$\\
\hline
\end{tabular}
\caption{Set of parameters for constructing the purified spin configurations from the $(a,b,c)$ and $(a_{1},b_{1},c_{1})$ phases according to the ansatz outlined in \ref{eq:3Q} for spin configurations shown in Fig. \ref{fig11}}
\label{table:tab8}
\end{table}


\begin{figure*}[htpb]
\centering
\includegraphics[width=\linewidth]{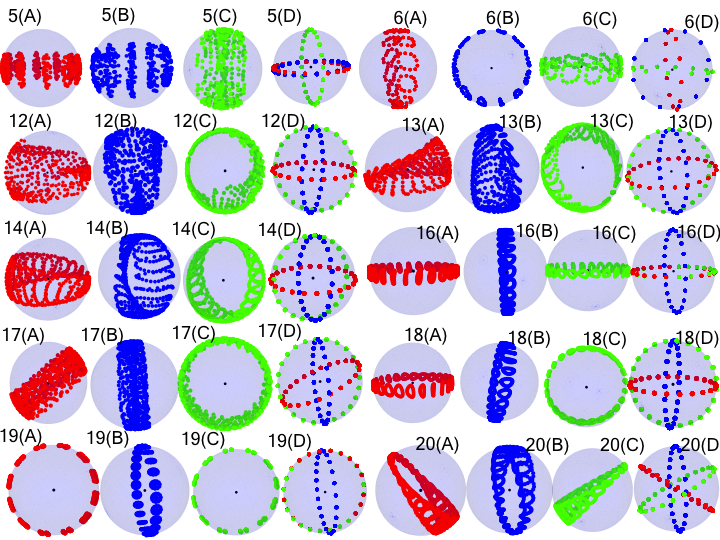} 
\caption{Common origin plots of the spin configurations from the $(a,b,c)$ and the $(a_{1},b_{1},c_{1})$ phases shown in Fig.4 in \cite{maintext}. (A)-(C) label common origin plots for spins on each of the three sublattices and (D) shows the spins in the "purified" configuration. (5):0.226, (6):0.28, (12):0.362, (13): 0.394, (14):0.485, (16):0.527, (17):0.551, (18):0.579, (19):0.596, (20):0.624.  }
\label{fig11}
\end{figure*}

\begin{table*}[ht]
\centering
\begin{tabular} {|p{0.4cm}|p{1.41cm}|p{1.41cm}|p{1.3cm}|p{2cm}|p{2cm}|p{2cm}|p{2cm}|p{2cm}|p{2cm}|}
 \hline\hline
$\#$ & $\hspace{3mm} \mathbf{q}_{1}(2\pi)$ & $\hspace{3mm} \mathbf{q}_{2}(2\pi)$ & $\hspace{3mm} \mathbf{q}_{3}(2\pi)$ &  $ \hspace{5mm} \mathbf{S}_{1}(\mathbf{q}_{1})$ & $\hspace{5mm} \mathbf{S}_{1}(\mathbf{q}_{2})$ & $\hspace{5mm} \mathbf{S}_{2}(\mathbf{q}_{2})$ & $\hspace{5mm} \mathbf{S}_{2}(\mathbf{q}_{3})$ & $\hspace{5mm} \mathbf{S}_{3}(\mathbf{q}_{1})$ & $\hspace{5mm} \mathbf{S}_{3}(\mathbf{q}_{3})$      \\ 
\hline\hline
$(1)$ & $\left(\begin{array}{c}0.19\\-0.13\end{array}\right)$ & $\left (\begin{array}{c} 0.02 \\ -0.22 \end{array} \right)$ & $\left (\begin{array}{c} 0.21 \\ 0.09 \end{array} \right)$ & $\left (\begin{array}{c} 0.33e^{i 1.4} \\ 0.3 e^{-i 0.3 } \\ 0.16 e^{i 3.1} \end{array} \right)$ & $\left (\begin{array}{c} 0.32 e^{-i 0.8 } \\ 0.33 e^{-i 2.4} \\ 0.11 e^{i 1.6} \end{array} \right)$ & $\left (\begin{array}{c} 0.36 e^{-i 0.6} \\ 0.37 e^{-i 2.3} \\ 0.13 e^{i 1.7} \end{array} \right)$ & $\left (\begin{array}{c} 0.09e^{-i 2.8} \\ 0.15 e^{-i 3.1} \\ 0.4e^{-i 3} \end{array} \right)$ & $\left (\begin{array}{c} 0.37e^{i 1.2} \\ 0.34 e^{-i 0.4} \\ 0.18e^{i 3} \end{array} \right)$ & $\left (\begin{array}{c} 0.09e^{i2.7} \\ 0.15e^{i 3} \\ 0.4e^{-i 3.1} \end{array} \right)$  \\
\hline\hline
$(2)$ & $\left(\begin{array}{c}0.15\\-0.1\end{array}\right)$ & $\left (\begin{array}{c} 0.02 \\ -0.18 \end{array} \right)$ & $\left (\begin{array}{c} 0.17 \\ 0.07 \end{array} \right)$ & $\left (\begin{array}{c} 0.29e^{i 1.1} \\ 0.15 e^{-i 2.3 } \\ 0.33 e^{-i 0.5} \end{array} \right)$ & $\left (\begin{array}{c} 0.29 e^{i 1.9 } \\ 0.15 e^{-i 1.5} \\ 0.32 e^{i 0.3} \end{array} \right)$ & $\left (\begin{array}{c} 0.33 e^{i 2} \\ 0.17 e^{-i 1.4} \\ 0.37 e^{i 0.4} \end{array} \right)$ & $\left (\begin{array}{c} 0.19e^{i 2} \\ 0.39 e^{2i} \\ 0.04e^{i 2} \end{array} \right)$ & $\left (\begin{array}{c} 0.33e^{i} \\ 0.16 e^{-i 2.4} \\ 0.37e^{-i 0.6} \end{array} \right)$ & $\left (\begin{array}{c} 0.19e^{ i1.9} \\ 0.39e^{i 1.9} \\ 0.04e^{i 1.9} \end{array} \right)$  \\
\hline\hline
$(3)$ & $\left(\begin{array}{c}{\small 0.02}\\-0.13\end{array}\right)$ & $\left (\begin{array}{c} -0.12 \\ 0.05 \end{array} \right)$ & $\left (\begin{array}{c} 0.1 \\ 0.08 \end{array} \right)$ & $\left (\begin{array}{c} 0.06e^{-i 0.6} \\ 0.03 e^{-i 0.6 } \\ 0.47 e^{i 2.5} \end{array} \right)$ & $\left (\begin{array}{c} 0.29 e^{i 2.6 } \\ 0.36 e^{-i 0.5} \\ 0.01 e^{i 2.6} \end{array} \right)$ & $\left (\begin{array}{c} 0.36 e^{-i 3} \\ 0.29 e^{-i 3} \\ 0.06 e^{-i 3} \end{array} \right)$ & $\left (\begin{array}{c} 0.06e^{-i 0.5} \\ 0.03 e^{-i 0.5} \\ 0.46e^{i 2.6} \end{array} \right)$ & $\left (\begin{array}{c} 0.3e^{i 2.5} \\ 0.37 e^{-i 0.6} \\ 0.01e^{i 2.5} \end{array} \right)$ & $\left (\begin{array}{c} 0.36e^{-i2.9} \\ 0.29e^{-i 2.9} \\ 0.06e^{-i 2.9} \end{array} \right)$  \\

\hline 
\end{tabular}
\caption{Ordering wave vectors and spin F.T. for reconstructing orders (1) at $n=0.115$, (2) at $n=0.146$ and (3) at $n=0.181$ occurring in the phase diagram in Fig.4 in \cite{maintext}. From left to right: order number corresponding to labeling in Fig. 4 in \cite{maintext}, Ordering wave vectors $\mathbf{q}_{1,2,3}$ for the $3\mathbf{Q}(ab,bc,ca)$ state, spin F.T. at each wave vector. An approximate and un-normalized spin order can be constructed using the information provided above using the recipe provided in text (see Eq.\ref{eq:s3q})}
\label{table:tab7}
\end{table*}


\begin{thebibliography}{99}
\bibitem{fn:bcaveraging}If, for that filling, some reciprocal vectors must
be partially occupied or ambiguous, we use averaging and
symmetrization procedures, interpolating between fillings
and ensuring the result always has the full symmetry of the lattice.
\bibitem{fn:statelabel}It does not specify whether the coefficients of the different wavevectors are equal or unequal, nor does it specific particular phase relationships between the amplitudes of the same wavevector appearing in different sublattices. These could distinguish two phases we denote with the same label.
\end{thebibliography}
\end{document}